\documentclass[9pt,english]{extarticle}
\usepackage[hang]{footmisc}

\usepackage{amsmath} 
\usepackage{epstopdf}
\usepackage{graphicx} 
\usepackage{balance} 

\usepackage{epstopdf}

\usepackage{multirow} 
\usepackage{array}

\usepackage{wrapfig} 
\usepackage[utf8]{inputenc} 
\usepackage{clrscode} 
\usepackage{listings} 
\usepackage{booktabs}

\usepackage{subfigure} 
\usepackage{hyperref}
\usepackage{gensymb}
\usepackage{tikz}
\usepackage{verbatim}
\usepackage{scalefnt}
\usetikzlibrary{arrows}
\usetikzlibrary{positioning}
\usetikzlibrary{decorations.markings}
\usetikzlibrary{decorations.pathreplacing}

\usepackage[utf8]{inputenc}
\usepackage{amsmath}
\usepackage{amsfonts}
\usepackage{amssymb}
\usepackage{amsthm}
\usepackage{graphicx}
\usepackage{color}
\usepackage[ruled,linesnumbered]{algorithm2e}
\usepackage{caption}
\usepackage{url}
\usepackage{booktabs}

\usepackage{palatino}

\newcommand{\ra}[1]{\renewcommand{\arraystretch}{#1}}

\DeclareMathAlphabet{\mathpzc}{OT1}{pzc}{m}{it}

\newcommand{\ignore}[1]{}

\usepackage[left=2cm,right=2cm,top=2cm,bottom=2cm]{geometry}

\usepackage{palatino}

\newcommand{\g}{\ensuremath{G}}
\newcommand{\gstar}{\ensuremath{G^*}}
\newcommand{\hstar}{\ensuremath{H^*}}
\newcommand{\gd}{Dij-G}
\newcommand{\gstard}{Dij-G\ensuremath{^*}}
\newcommand{\hstard}{Dij-H\ensuremath{^*}}
\newcommand{\piandpj}{$P_i$ and $P_j$}
\newcommand{\robsij}{\ensuremath{R_{i,j}}}
\newcommand{\rij}{\ensuremath{R_{i,j}}}
\newcommand{\dij}{\ensuremath{D_{i,j}}}
\newcommand{\wij}{\ensuremath{\mathcal{W}_{i,j}}}

\DeclareFontFamily{OT1}{pzc}{}
\DeclareFontShape{OT1}{pzc}{m}{it}{<-> s * [1.10] pzcmi7t}{}
\DeclareMathAlphabet{\mathpzc}{OT1}{pzc}{m}{it}

\usepackage{tikz}
\usepackage{verbatim}
\usepackage{scalefnt}
\usetikzlibrary{arrows}
\usetikzlibrary{positioning}
\usetikzlibrary{decorations.markings}
\usetikzlibrary{decorations.pathreplacing}

\begin{document}

\title{\Huge{Towards Knowledge-Enriched Path Computation}}

\author{
  Georgios Skoumas\\
  School of Electrical and Computer Engineering, NTUA \\
  \texttt{gskoumas@dblab.ece.ntua.gr}
  \and
  Klaus Arthur Schmid\\
  Department of Computer Science, LMU\\
  \texttt{schmid@dbs.ifi.lmu.de}
  \and
  Gregor Joss\'e\\
  Department of Computer Science, LMU\\
  \texttt{josse@dbs.ifi.lmu.de}
  \and
  Andreas~Z\"{u}fle\\
  Department of Computer Science, LMU\\
  \texttt{zuefle@dbs.ifi.lmu.de}
  \and
  Mario A. Nascimento\\
  Department of Computing Science, UofA\\
  \texttt{nascimento@ualberta.ca}
  \and
  Matthias Renz\\
  Department of Computer Science, LMU\\
  \texttt{renz@dbs.ifi.lmu.de}  
  \and
  Dieter Pfoser\\
  Department of Geography and Geoinformation Science, GMU \\
  \texttt{dpfoser@gmu.edu}
}


%
%
%

\maketitle

\begin{abstract}
Directions and paths, as commonly provided by navigation systems, are
usually derived considering absolute metrics, e.g., finding
the shortest path within an underlying road network.  With the aid of
crowdsourced geospatial data we aim at obtaining paths that do not only
minimize distance but also lead through more popular areas using
knowledge generated by users. We extract spatial relations such as ``nearby'' or ``next to'' from travel blogs,
that define closeness between pairs of points of interest (PoIs) and quantify
each of these relations using a probabilistic model. Subsequently, we
create a relationship graph where each node corresponds to a PoI and
each edge describes the spatial connection between the respective PoIs.
Using Bayesian inference we obtain a probabilistic measure of spatial closeness according to the crowd. 
Applying this measure to the corresponding road network, we obtain an altered cost function which does not exclusively rely on distance, and enriches an actual road networks taking  
crowdsourced spatial relations into account. 
Finally, we propose two routing algorithms on the enriched road networks. 
To evaluate our approach, we use Flickr photo data as a ground truth for popularity. 
Our experimental results -- based on real world datasets -- show that the paths computed w.r.t.\ our alternative cost function yield competitive solutions in terms of path
length while also providing more ``popular'' paths, making routing easier and more informative for the user.

\end{abstract}

\section{Introduction}
\label{sec:introduction}
\vspace{6pt}

User-contributed content has benefited many scientific disciplines by providing
a wealth of new data. Technological progress, especially smart phones and GPS
receivers, has facilitated contributing to the plethora of available
information. OpenStreetMap\footnote{\url{https://www.openstreetmap.org/}}
constitutes the standard example and reference in the area of
volunteered geographic information. Authoring geospatial information
typically implies coordinate-based, \emph{quantitative data}.
Contributing quantitative data requires specialized applications (often part of social media platforms)
and/or specialized knowledge, as is the case with OpenStreetMap (OSM).


The broad mass of users contributing content, however, are much
more comfortable using \emph{qualitative information}. People
typically do not use geographical coordinates to describe their spatial
motion, for instance when traveling or roaming. Instead, they use
qualitative information in the form of toponyms (landmarks) and
spatial relationships (``near'', ``next to'', ``north of'', etc.).
Hence, there is an abundance of geospatial information (freely)
available on the Internet, e.g., in travel blogs,
largely unused. In contrast to quantitative information, which is
mathematically measurable (although sometimes flawed by measurement
errors), qualitative information is based on personal cognition.
Therefore, accumulated and processed qualitative information
may better represent human way of thinking.

This is of particular interest when considering the ``routing
problem'' (equivalent to ``path finding''). Traditional routing
queries use directions from systems that only take the structure
of the underlying road network into account. In human interaction
such information is usually enhanced with qualitative information
(e.g.\ ``the street next to the church'', ``the bridge north of the
Eiffel tower''). Combining traditional routing algorithms with
crowdsourced -- thus, more commonly used -- geospatial references
we aim to more properly represent human perception while keeping
it mathematically measurable.
In this work, we enrich a road network with information
about spatial relations between pairs of Points of Interest (PoI)
extracted from textual travel-blog data. Using these relations,
we obtain routes that are easier to interpret and to
follow, much rather resembling a route that a person would
provide.

\begin{figure}[t]
  \begin{center}
  \includegraphics[width=0.47\textwidth]{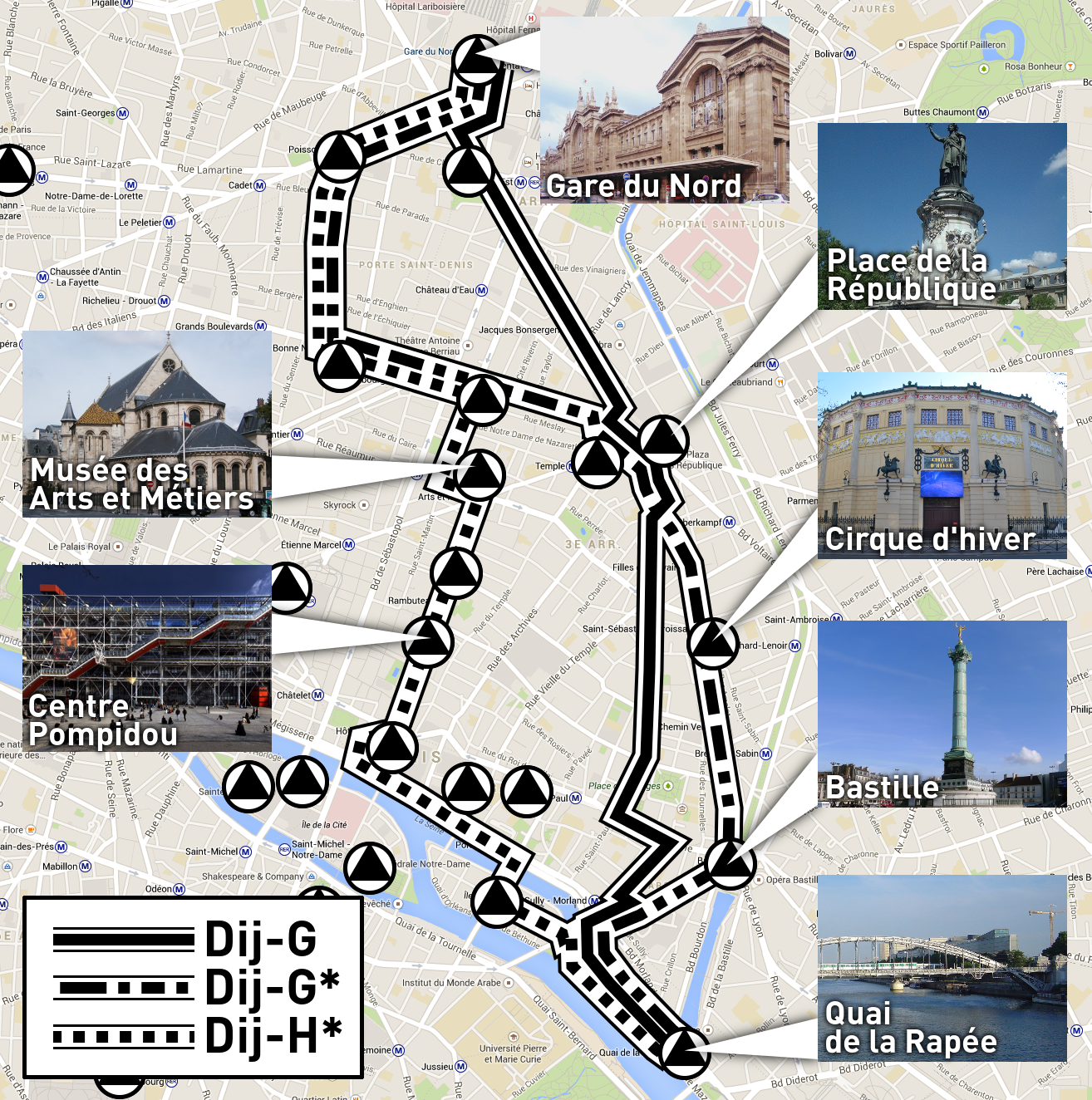}
  \caption{Shortest (continuous) and alternative paths (dot dashed and dotted) alongside
  PoIs in the city of Paris. This result is an output of the algorithms presented in this paper. \vspace{-10pt}}
\label{fig:intuition}
  \vspace{-10pt}
\end{center}
\end{figure}

As an example, consider the routing scenario in
Figure~\ref{fig:intuition} which is set in the city of Paris,
France. The continuous line represents the conventional shortest
path from starting point ``Gare du Nord'' to the target
at ``Quai de la Rap\'{e}e'' \textendash while the dot dashed and
dotted lines represent alternative paths computed by the
algorithms introduced in this paper. The triangles in this
example denote touristic landmarks and sights. For instance,
the dot dashed path on the bottom right passing recognizable
locations such as ``Place de la R\'{e}publique'', ``Cirque
d'hiver'' and ``la Bastille'', as proposed by our algorithms,
is considerably easier to describe and follow, and might yield
more interesting sights for tourists than the shortest path.

The challenge of this work is to extract crowdsourced information
from textual data and incorporate it an existing road network.
This enriched road network is subsequently used to provide
paths between a given start and target that satisfy the claim of
higher popularity (which is formally introduced in Section
\ref{sec:relgraphs}), while only incurring a minor additional
spatial distance. In addition to this main application, we note
that our techniques can furthermore be used to automatically
provide interesting tourist routes in any place where information
about PoIs is publicly available.

The transition from textual information to routing in networks is not at all straightforward,
therefore we employ and develop various methods from different angles of computing science:
\begin{itemize}
  \item We begin by mining information from texts, employing Natural Language Processing methods in
  order to determine spatial entities and relations between them (cf.\ Section~\ref{sec:contribution}).
  
  \item Due to the inherent uncertainty of crowdsourced data, we employ 
  probability distributions to quantitatively model spatial relations mined from the text (cf. Section~\ref{sec:modeling}).

  \item We propose a Bayesian inference-based transition from the probabilistically modeled spatial relations 
   to confidence measurements about the quality of spatial relation extracted from text (cf. Section~\ref{subsec:rel2weight}).
  
  \item We define a new cost criterion which is used to enrich an underlying road network with the aforementioned confidence measurements (cf. Section~\ref{subsec:enrich}).
  
  \item Finally, we propose two algorithms which use the enriched road network to compute actual paths (cf.\ Section~\ref{subsec:algorithms}).
\end{itemize}

\section{Related Work}
 
\label{sec:related_work}
\vspace{6pt}

Research areas relevant to this work include: $(i)$ qualitative routing and $(ii)$
mining of semantic information from moving object trajectories and trajectory enrichment with
extracted semantic information. In what follows, we discuss previous work in both of these areas.


While finding shortest paths in road networks is a thoroughly explored
research area, qualitative routing has hardly been explored. Nevertheless, providing meaningful
routing directions in road networks is a research topic of great importance. In various real world
scenarios, the shortest path may not be the ideal choice for providing directions
in written or spoken form, for instance when in an unfamiliar neighborhood, or in cases of
emergency. Rather, it is often more preferable to offer ``simple'' directions that are easy to
memorize, explain, understand and follow. However, there exist cases where the simplest route is
considerably longer than the shortest. The authors in \cite{Sacharidis}, \cite{Kuhn:SIT} and
\cite{Westphal:QualitativeRouting} try to tackle the problem of efficient routing by using cost 
functions that trade off between minimizing the length of a provided path while also minimizing the 
number of turns, e.g., number of instructions, of the provided path. The major shortcoming of these approaches is that they focus 
almost exclusively on road network data without taking into account any kind of
qualitative information, i.e.\ information coming from the user.
Opposed to that, we try to approach the problem of efficient routing by integrating spatial knowledge coming from the crowd thus 
enriching an actual road network.

The importance of landmark based routing from human cognition point of view, has been analyzed on a theoretical basis, e.g., without an real routing application, in \cite{Raubal:2002:EWI:646933.759822} and 
\cite{Raubal:GIS}. The discovery of semantic places through the analysis of raw trajectory
data has been investigated thoroughly over the course of the last years.
The authors in \cite{Lv:PersonalSemPlaceMining}, \cite{Yan:SemanticAnnotation} and \cite{Palma:ClusteringApprForPlaceMining} 
provide solutions for the semantic place recognition problem and categorize the extracted PoIs into pre-defined types. 
Moreover, the concept of ``semantic behavior'' has recently been introduced.
This refers to the use of semantic abstractions of the raw mobility data, 
including not only geometric patterns but also knowledge extracted jointly 
from the mobility data as well as the underlying geographic and application 
domains in order to understand the actual behavior of moving users.
Several approaches like \cite{Alvares:EnrichTraj}, 
\cite{EPFL:SemanticTraj1}, \cite{EPFL:SemanticTraj4}, \cite{EPFL:SemanticTraj2}, \cite{EPFL:SemanticTraj4} and \cite{EPFL:SemanticTraj6}
have been introduced the last decade. The core contribution of these articles lies in the development of a semantic approach that progressively transforms the raw mobility data into 
semantic trajectories enriched with PoIs, segmentations and annotations. Moreover, a recent work, \cite{Feldman:iDiary}, can extract and transform the aforementioned semantic
information into a text description in the form of a diary. 
Finally, the authors in \cite{DBLP:Yahoo} propose an approach for route recommendation based on a user opinion mining platform 
about trajectories, providing longer but more pleasant routes. The major drawback of these approaches is that they do not integrate the extracted semantic information into the road network. 
Instead, they use the extracted information only on specific trajectories. 
In our contribution, we analyze crowdsourced data in order to extract semantic spatial information and integrate
it into an actual road network. This will enable us to provide routes that are near-optimal
w.r.t. distance while spatially more popular according to the crowd.

\section{Spatial Relation Extraction}
\label{sec:contribution}
\vspace{6pt}

%
%
In this work, we choose travel blogs as a rich potential source for (crowdsourced) geo-spatial data.
This selection is based on the fact that people tend to describe their experiences in relation
 to their trips and places they have visited, which results in ``spatial'' narratives.
To gather such data, we use classical Web crawling techniques and compile a database consisting
of 120,000 texts, obtained from travel blogs\footnote{\url{http://www.travelblog.com/} \\  \url{http://www.traveljournal.com/} \\ \url{http://www.travelpod.com/}}.

Obtaining qualitative spatial relations from text involves the detection of $(i)$
PoIs (or toponyms) and $(ii)$ spatial relationships linking the PoIs.
The employed approach involves geoparsing, i.e. the detection of
candidate phrases, and geocoding, i.e., linking the phrases to actual coordinate information.

Using the Natural Language Processing Toolkit (NLTK) (cf. \cite{nltk}),
a leading platform for analyzing raw natural language data,
we managed to extract 500,000 PoIs from the text corpus.
For the geocoding of the PoIs, we rely on the GeoNames\footnote{\url{http://www.geonames.org/}}
geographical gazetteer data, which contains over ten million PoI names worldwide and their coordinates.
This procedure associates (whenever possible) PoIs found in the travel blogs with geo coordinates.
Using the GeoNames gazetteer we were able to geocode about 480,000 out of the 500,000 extracted PoIs.

Having identified and geocoded the spatial objects, the next step is the extraction of qualitative spatial relationships.
The extraction of spatial relations between entities in text is a hard Natural Language Processing (NLP) problem,
especially when applied to a noisy crowdsourced dataset. 
We address this NLP challenge by implementing a spatial relation extraction algorithm based on NLTK \cite{nltk} components in combination with predefined strings and syntactical patterns.
More specifically, we define a set of language expressions that are typically used to express a spatial relation in combination with a set of syntactical rules.
The use of both syntactical and string matching reduces the number of false positives considerably. As an
example, consider the following phrase. \textit{``\textbf{Deutsche Bank} invested 10 million dollars \textbf{in} \textbf{Brazil}''}. Here, a simple string
matching solution would return a triplet of the form (Deutsche Bank, in, Brazil), which is a false positive. In our approach, the use of predefined syntactical
patterns avoids this kind of mistakes.


Algorithm~\ref{alg:relext} describes the architecture of the proposed information extraction system. Initially, syntactical and string patterns are loaded (Step 2).
The raw text document is segmented into sentences  (Step 4). Each sentence is further subdivided (tokenized) into words and tagged as part-of-speech (Steps 6-7).
Continuing, name entities (PoIs) are identified (Step 8). We look for relations between specified types of named entities,
which in NLTK are \emph{Organizations, Locations, Facilities and Geo-Political Entities (GPEs)}.
Next, in case there are two or more name entities in the sentence, we check if any of the predefined syntactical patterns applies between the recognized name entities pairs (Step 13).
If it exists, we then use regular expressions to determine the specific spatial relation instance from our predefined spatial relation pattern list for this case.
If there is a string pattern match we record the extracted triplet (Steps 15-16).
Thus, the search for spatial relations in texts results into triplets of the form ($P_i$, $R^k$, $P_j$), 

\IncMargin{1em}
\begin{algorithm}[!htb]
\DontPrintSemicolon
 \KwIn{A database of texts $\mathcal{T}$}
 \KwOut{A set of triplets $\mathcal{O}= (P_i, R^k, P_j)$ where $P_{i} \neq P_{j}$}
 \BlankLine
 \Begin{
  Load syntactical $\mathcal{A}$ and string $\mathcal{B}$ patterns

  \ForEach{\text{text} $t \in \mathcal{T}$}{
   Extract sentences from $t$ into set $\mathcal{S}$

    \ForEach{sentence $s \in \mathcal{S}$}{
      Token $s$ using NLTK \;
      PosTag $s$ using NLTK\;
      Identify name entities using NLTK

      \If{two or more name entities in $s$}{
		Extract pairs in $\mathcal{P}$
		
		\ForEach{$p \in \mathcal{P}$}{
	   		$p_{\mathcal{A}} \leftarrow$ Extract syntactical pattern of $p$
	    	
	    	\If{$p_{\mathcal{A}} \in \mathcal{A}$}{	
				$p_{\mathcal{B}} \leftarrow$ Extract string pattern of $p$				

				\If{$ p_{\mathcal{B}} \in \mathcal{B}$}{	
		    		$\mathcal{O}.\proc{PushTriplet}(p(1),p_{\mathcal{B}},p(2))$
				}
	      	}
	  	}
      }
    }
  }

\BlankLine

\Return{$\mathcal{O}$} \;
\BlankLine
}%
\caption{Spatial Relation Extraction}
\label{alg:relext}
\end{algorithm}
\DecMargin{1em}

\begin{figure}[h!]
\centering
\includegraphics[width=0.58\textwidth]{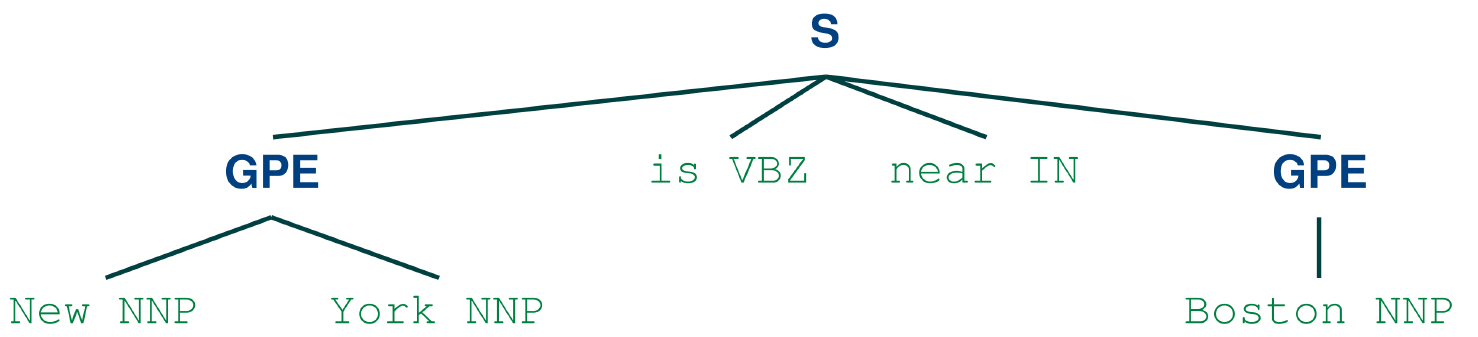}
\caption{Example of a parsed sentence syntactic tree.}
\label{fig:parsetree}
\end{figure}

An example is shown in Figure~\ref{fig:parsetree}, where a sentence is analyzed as explained and two named entities are identified as GPEs.
We first check the syntax and make sure that the pattern
\textit{``GPE - 3rd person verbal phrase (VBZ) - preposition/subordinating conjunction (IN) - GPE''} exists in our set of predefined spatial relation patterns.
Performing string matching on the intermediate chunks (``near'') results in the triplet \textit{(New York, Near, Boston)}.
Applying Algorithm~\ref{alg:relext}, we extracted 440,000 triplets from our 120,000 travel blog text corpus.

\begin{figure}[h!]
\centering
\includegraphics[width=0.6\textwidth]{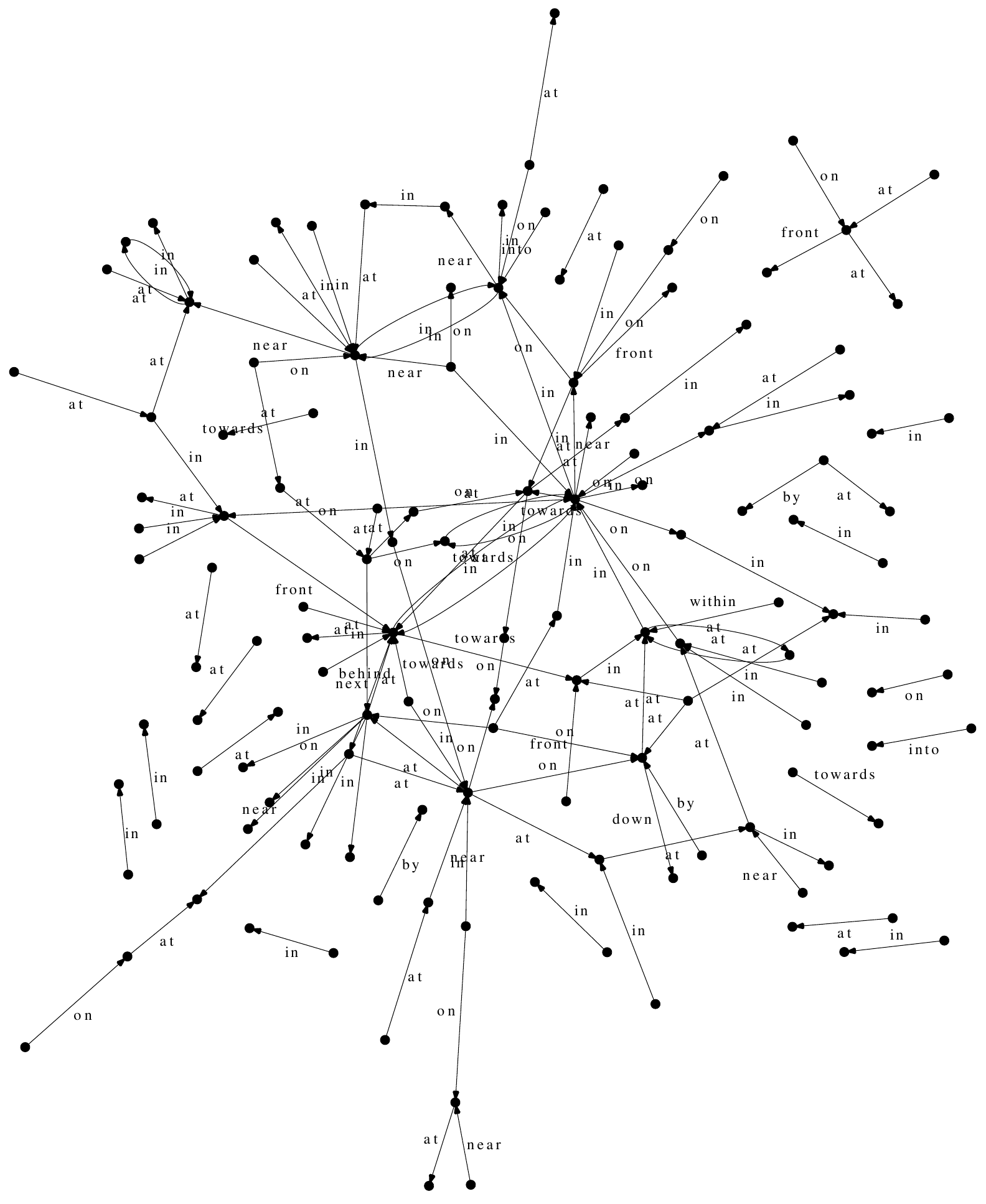}
\caption{Sample of a spatial relation graph for London.}
\label{fig:relgraph1}
\end{figure}

Figure~\ref{fig:relgraph1} shows a \emph{Spatial Relationship Graph}, i.e., a spatial graph in which nodes represent PoIs and edges label spatial relationships existing between them.
The graph visualizes a small sample of spatial relationship data collected for the city of London.
We have collected data for the regions of New York, Paris and London which will be our datasets during the experimental evaluation of the proposed approach.
Here, we should point out that for the scope of this work, i.e. a combination of short and enriched routes, we only consider distance and topological relations
that denote closeness (near, close, next to, at, in etc). The use of relations that denote direction, e.g. north, south, east etc., or remoteness, e.g.
away from, far etc., is an open direction for future work.

%
%
%

\section{Modeling Spatial Relations}
\label{sec:modeling}
\vspace{6pt}
In this section we describe the probabilistic modeling we follow in order to quantify the qualitative spatial relations between PoIs.
Key ingredients of our system are methods that train probabilistic models, which is required due to the inherent uncertainty of crowdsourced data.
Our discussion below includes a description of the features we used to model spatial relations and a short analysis
of the modeling approach used in \cite{skoumas}, i.e., the probabilistic mixture models we employ for the quantitative representation of spatial relations
and a greedy learning algorithm for model parameter estimation. These spatial relations models are necessary
to assess the quality of spatial relations extracted from text. The spatial relations and their corresponding
quality information will be used in Section \ref{subsec:enrich} for the enrichment of the underlying spatial network.
This enrichment will be used to find paths which favor popular parts of the road network in Section \ref{subsec:algorithms}.


\subsection{Spatial Features}
\label{subsection:spatfeats}
\vspace{6pt}
In this work, we model a spatial relation between two PoIs \piandpj \ in terms of \emph{distance} and \emph{orientation} as presented in \cite{skoumas}.
Therefore, we extract occurrences of a spatial relation (such as ``near'') from travel blogs as described in Section~\ref{sec:contribution}.
For each occurrence, we create a two-dimensional spatial feature
vector $D = (D_d, D_o)^{\intercal}$ where $D_d$ denotes
the distance and $D_o$ denotes the orientation between $P_i$ and $P_j$.
This way, we end up with a set of two-dimensional feature vectors $\mathcal{D}_\text{rel}=\{D_1, D_2, \dots, D_n\}$ for each spatial relation. An example of the feature extraction procedure is illustrated
in Figure~\ref{fig:featext}  (cf. \cite{skoumas}), where four instances of spatial relation \textit{near} create the respective
set of spatial feature vectors $\mathcal{D}_{\text{near}}= \{(D_{d1}, D_{o1})^{\intercal}, (D_{d2}, D_{o2})^{\intercal}, (D_{d3}, D_{o3})^{\intercal}, (D_{d4}, D_{o4})^{\intercal} \}$.

\tikzset{
  big arrow/.style={
    decoration={markings,mark=at position 1 with {\arrow[scale=1.5,#1]{>}}},
    postaction={decorate},
    shorten >=0.2pt},
  big arrow/.default=black}

\begin{figure}[h!]
\scalefont{0.5}
\begin{center}
\begin{tikzpicture}[scale=1.0]
    \draw [<->,thick] (0,7) node (yaxis) [above] {$y$}
        |- (10,0) node (xaxis) [right] {$x$};
    \coordinate[label=200:$A$]  (a) at (1.0,1.0);
    \coordinate[label=45:$B$]   (b) at (2.0,2.2);
    \coordinate[label=290:$C$]  (c) at (1.5,4.2);
    \coordinate[label=290:$D$]  (d) at (3.0,5.0);
    \coordinate   (e) at (6.5,5.5);
    \coordinate[label=0:$F$]   (f) at (7.5,4.8);
    \coordinate[label=290:$G$]  (g) at (8.0,1.0);
    \coordinate[label=45:$H$]  (h) at (6.5,3.3);

    \coordinate[label=290:$E$] (k) at (6.5,6.0);

    \fill[black] (a) circle (1.2pt);
    \fill[black] (b) circle (1.2pt);
    \fill[black] (c) circle (1.2pt);
    \fill[black] (d) circle (1.2pt);
    \fill[black] (e) circle (1.2pt);
    \fill[black] (f) circle (1.2pt);
    \fill[black] (g) circle (1.2pt);
    \fill[black] (h) circle (1.2pt);

    \draw[black, big arrow] (a)--(b);
    \draw[black, big arrow] (d)--(c);
    \draw[black, big arrow] (e)--(f);
    \draw[black, big arrow] (g)--(h);

    \draw [dashed] (0.5,1)--(3,1);
    \draw [dashed] (1,2.5)--(1,0.5);

    \draw [dashed] (3,6)--(3,4.5);
    \draw [dashed] (2.5,5)--(4.5,5);

    \draw [dashed] (6.0,5.5)--(8,5.5);
    \draw [dashed] (6.5,5)--(6.5,6.5);

    \draw [dashed] (8,0.5)--(8,2.5);
    \draw [dashed] (7.5,1.0)--(9.7,1.0);

    \draw [black!50!black,thick](a) +(0:.3cm) arc (0:37:.4cm);
    \draw [color=black](a)+(15:0.65) node[rotate=0] {$D_{o1}$};

    \draw [black!50!black,thick](d) +(0:.2cm) arc (0:210:.2cm);
    \draw [color=black](d)+(135:0.55) node[rotate=0] {$D_{o2}$};

    \draw [black!50!black,thick](e) +(0:.2cm) arc (0:330:.2cm);
    \draw [color=black](e)+(145:0.55) node[rotate=0] {$D_{o3}$};

    \draw [black!50!black,thick](g) +(0:.2cm) arc (0:120:.2cm);
    \draw [color=black](g)+(35:0.55) node[rotate=0] {$D_{o4}$};

    \draw [decorate,decoration={brace,amplitude=5pt},xshift=-4pt,yshift=0pt]
    (a)--(b) node [black,midway,xshift=-0.15cm,yshift=+0.3cm] {$D_{d1}$};

    \draw [decorate,decoration={brace,amplitude=5pt},xshift=-4pt,yshift=0pt]
    (d)--(c) node [black,midway,xshift=+0.32cm,yshift=-0.32cm] {$D_{d2}$};

    \draw [decorate,decoration={brace,amplitude=5pt},xshift=-4pt,yshift=0pt]
    (f)--(e) node [black,midway,xshift=+0.0cm,yshift=-0.33cm] {$D_{d3}$};

    \draw [decorate,decoration={brace,amplitude=5pt},xshift=-4pt,yshift=0pt]
    (g)--(h) node [black,midway,xshift=-0.3cm,yshift=-0.2cm] {$D_{d4}$};
\end{tikzpicture}
\caption{Distance and orientation feature extraction procedure - B is near A, C is near D, F is near E and H is near G.} \label{fig:featext}
\end{center}

\end{figure}
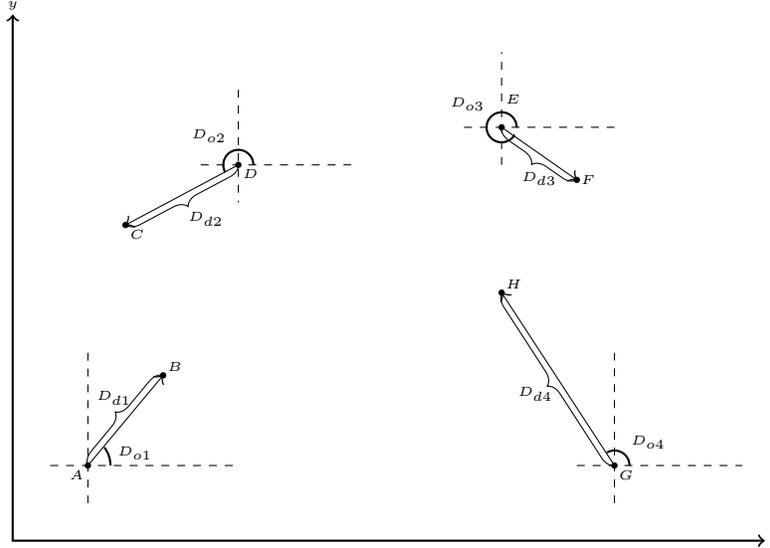

\subsection{Probabilistic Models}
\vspace{6pt}
Section \ref{subsection:spatfeats} provides us, for each spatial relation such as ``near'' and ``into'', with a set of distances and orientations of their instances.
Using these instances, we employ Gaussian Mixture Models (GMMs) which have been extensively
 used in many classification and general machine learning problems (\cite{Bishop}).
  They are very well known for $(i)$ their formality, as they build on the formal
  probability theory, $(ii)$ their practicality, as they have been implemented several
   times in practice, $(iii)$ their generality, as they are capable of handling many
   different types of uncertainty, and $(iv)$ their effectiveness. 

In general, a GMM is a weighted sum of $M$-component Gaussian densities as $p(d|\lambda) = \sum_{i=1}^{M} w_{i} g(d;\mu_{i},\Sigma_{i})$
where $d$ is a $l$-dimensional data vector (in our case $l=2$), $w_i$ are the mixture weights, and $g(d;\mu_i, \Sigma_i )$ is a Gaussian density function
with mean vector $\mu_i \in \mathbb{R}^{l} $ and covariance matrix $\Sigma_i \in \mathbb{R}^{l\times l}$.
To fully characterize the probability density function $p(d|\lambda)$, one requires the mean vectors, the covariance matrices and the mixture weights. These parameters are collectively represented 
in $\lambda = \{w_i, \mu_i, \Sigma_i\}$ for $i= 1, \dots, M$.

In our setting, each spatial relation is modeled under a probabilistic framework by a 2-dimensional GMM, trained on each relation's spatial feature vectors,
as discussed in Section~\ref{subsection:spatfeats}.
For the parameter estimation of each Gaussian component of each GMM, we use Expectation Maximization (EM) (\cite{Dempster}).
EM enables us to update the parameters of a given M-component mixture with respect to a feature vector
set $\mathcal{D}_\text{rel} = \{D_1 , \dots, D_n \}$ with $1 \leq j \leq n$ and all $D_j \in \mathbb{R}^{l}$, such that the log-likelihood given
by Equation~\ref{eq:loglike}, increases with each re-estimation step, i.e., EM re-estimates model parameters $\lambda$ until convergence. A formal
analysis of the parameter re-estimation formula for each EM step is given in the Appendix. 

\begin{equation}\label{eq:loglike}
\mathcal{L} = \sum_{j=1}^{n} \log(p(D_j|\lambda))
\end{equation}


A main issue in probabilistic modeling with probability mixtures is that a predefined number of components is neither a dynamic nor an efficient and robust approach.
The optimal number of components should be decided based on each dataset. We employ a greedy learning approach to dynamically estimate the number of
components in a GMM. This approach builds the mixture component in an efficient way by starting from an one-component GMM---whose parameters are trivially computed by using EM---
and then employing the following two basic steps until a stopping criterion is met:
\vspace{-5pt}

\begin{enumerate}
\item Insert a new component in the mixture
\item Apply EM until the log-likelihood $\mathcal{L}$ or the parameters of the GMM converge
\end{enumerate}

The stopping criterion can either be a maximum pre-selected number of components, or it can be any other model selection criterion.
In our case, the algorithm stops if the maximum number of components is reached,
or if the new model's log-likelihood $\mathcal{L}+1$ is less or equal to the log-likelihood $\mathcal{L}$ of the previous model,
after introducing a new component. A detailed description of the optimized GMM training algorithm is given in \cite{Verbeek}.

\section{ROAD NETWORK ENRICHMENT}
\label{sec:relgraphs}
\vspace{10pt}

In this section we describe our approach to enrich an actual road network with crowdsourced spatial information.
Our discussion below includes a description of how we transform a relationship graph, as presented in Section~\ref{sec:contribution},
into a weighted graph, and how we use the edge weights of the weighted graph in order to modify the
edge costs of a real road network.


%

\vspace{10pt}
\subsection{From Relationship to Weighted Graphs}
\label{subsec:rel2weight}
\vspace{10pt}

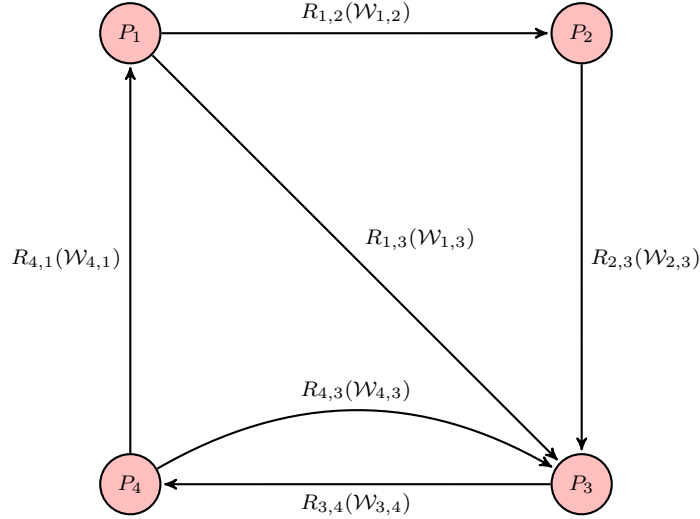
\begin{figure}[h!]
 \begin{center}
 \begin{tikzpicture}[scale=.8,every node/.style={minimum size=0.8cm},->,>=stealth',shorten >=1pt,auto,node distance=6.0cm,
  thick,main node/.style={circle,fill=red!25,draw,font=\sffamily\small\bfseries}, path node/.style={rectangle,fill=blue!20,draw,font=\sffamily\small\bfseries}, on grid]

  \node[main node, scale=1.0] (1) at (2.5,4.5) {$P_1$};
  \node[main node, scale=1.0] (2) [right of=1] {$P_2$};
  \node[main node, scale=1.0] (3) [below of=2] {$P_3$};
  \node[main node, scale=1.0] (4) [below of=1] {$P_4$};

  \path[every node/.style={font=\sffamily\small}]
    (1) 
        edge  node {$R_{1,2} (\mathcal{W}_{1,2})$} (2)
        edge  node {$R_{1,3} (\mathcal{W}_{1,3})$} (3)
    (2) edge node  {$R_{2,3} (\mathcal{W}_{2,3})$} (3)

    (3)  edge node {$R_{3,4} (\mathcal{W}_{3,4})$} (4)

     (4) edge node {$R_{4,1} (\mathcal{W}_{4,1})$} (1)
         edge [bend left] node[loop above] {$R_{4,3} (\mathcal{W}_{4,3})$} (3);
%
\end{tikzpicture}
\caption{Simple relationship graph. Nodes represent PoIs and each edge represents the set of relations
\robsij through which its adjacent nodes $P_i$ and $P_j$ are connected. Each of these sets is
mapped onto the closeness score \wij, turning the relationship into a weighted graph.}
\label{fig:relgraph2}
\end{center}
\end{figure}

 As presented in Section~\ref{sec:contribution}, the spatial relation extraction
 procedure results in a relationship graph between PoIs such as the relationship graph of Figure~\ref{fig:relgraph1}.
 A simpler example of such a graph is shown in Figure~\ref{fig:relgraph2}. In general,
 let $\mathcal{P} = \{ P_{1}, \dots, P_{m} \}$ denote the set of
 nodes representing the PoIs and let $\mathcal{R}  = \{ R^1, \dots, R^n\}$
 denote the pre-defined set of spatial closeness relations, represented by spatial NLP expressions
 like ``next to'' or ``close by''. Furthermore, let $\rij \subseteq \mathcal{R}$
 denote the set of relations observed (i.e. extracted from the text) between two distinct nodes \piandpj.
 Note that $R^k$ denotes an abstract relation, while $\rij$ denotes a set of occurrences of relations
 between a pair of nodes.
 Let $\dij$ denote the spatial feature vector (distance and orientation), 
between two distinct PoIs \piandpj \ (as presented in Section~\ref{subsection:spatfeats}).
 Finally, let $\mathcal{D} := \bigcup_{i\neq j \land \robsij \neq \emptyset}D_{i,j}$
 denote the set of all spatial feature vectors between all pairs of PoIs which have non-empty sets of relations.


%
We want to estimate the posterior probability of a class $R^k \in \robsij$
based on the spatial feature data $\dij$ between two PoIs \piandpj.
This is given by Equation~\ref{eq:bayes}.
$P(\dij|R^k)$ denotes the likelihood of \dij \ given relation $R^k$ based on the
trained GMM, while $P(R^k)$ denotes the prior probability of relation $R^k$ given only the
observed relations \robsij.
\vspace{-5pt}

 \begin{equation}\label{eq:bayes}
 P(R^k|\dij) 
    = \frac{P(\dij|R^k) P(R^k)}{\sum\limits_{l=1}^{n}  P(\dij|R^l) P(R^l)}
 \end{equation}

In a traditional classification problem the spatial relation $R^k$ between a
pair of PoIs would be classified to the spatial relation model
with the highest posterior.
In contrast to this approach, we consider each posterior probability $P(R^k|\dij)$
as a measure of confidence of the existence of relation $R^k$ between \piandpj.
Remember that all the relations we consider reflect terms of spatial closeness.
We combine all these posteriors into one measure which we refer to as \textit{closeness
score} \wij \ of the pair of PoIs \piandpj, defined in Equation~\ref{eq:spatconf}.

\begin{equation} \label{eq:spatconf}
  \wij =  \frac{1}{|\mathcal{R}|}  \cdot \sum\limits_{k=1}^{|\rij|} \frac{P(R^k|\dij)}{\max_k\{P(R^k|\mathcal{D})\}}
\end{equation}

Here, we sum all the posteriors $P(R^k|\dij)$ normalized by the maximum posterior
of each relation in the relationship graph and we normalize the summation
by the total number of spatial relations in the relationship graph. This is done for all
pairs $P_i,P_j$ where $\robsij \neq\emptyset$. We refer to these pairs as
\emph{close} since at least one of our relations reflecting closeness exists. As is illustrated in
Figure~\ref{fig:relgraph1}, assigning the respective weights \wij \ to the edges of the relationship
graph, we obtain a weighted graph. Note that $\wij\in[0,1]$ but typically $0<\wij\ll 1$.
In Section~\ref{sec:evaluation} the influence of $\wij$ on the results is examined, in particular,
different scalings are tested. In this weighted relationship graph, denoted by $H^*$, there exists a vertex for each PoI and an edge $(P_i,P_j)$ (equipped with weights
$\wij$ and Euclidean distances $d_{ij}$) for each pair of PoIs $P_i,P_j$ that are close in the above
sense ($\robsij \neq\emptyset$).



\vspace{10pt}\subsection{From Weighted Graphs to Road Network Enrichment}
\label{subsec:enrich}
\vspace{10pt}

Now that we have extracted and statistically condensed the crowdsourced data into a closeness score,
we need to apply the obtained closeness scores to the underlying network. We have investigated
several strategies and have decided upon a compromise between simplicity and effectiveness.

Let $G=(V,E,d)$ denote the graph representing the underlying road network,
i.e. the vertices $v\in V$ correspond to crossroads, dead ends, etc., the edges $e\in E \subseteq V \times V$
represent roads connecting vertices. Furthermore, let $d:E\rightarrow\mathbb{R}^+_0$ denote the function which maps every edge onto its distance.
We assume that $\mathcal{P} \subseteq V$, i.e. each PoI is also a vertex in the graph. This is only a minor
constraint since we can easily map each PoI to the nearest node of the graph or introduce pseudo-nodes.

Now, for each pair of spatially connected PoIs, $P_i,P_j$, we
compute the shortest path connecting \piandpj \ in $G$, which we denote by $p(i,j)$. We then
define a new cost function $c:E\rightarrow \mathbb{R}^+_0$ which modifies the previous cost $d(e)$ of an edge as follows:
\[c(e):= d(e)\cdot \prod_{e \in p(i,j)}(1- \alpha \wij),\]
where $e\in p(i,j)$ iff $e$ is an edge within the shortest path from $P_i$ to $P_j$ and where
$\alpha\in [0,1]$ is a weight scaling factor to control the balance between the spatial distance
$d(e)$ and the modification caused by the closeness score \wij. In the case of $\alpha=0$, we obtain the unadapted edge weight $c(e)=d(e)$.
Summarizing, the more shortest paths between PoI pairs run through $e$, the lower its adjusted cost $c(e)$ is.
The reason for enriching the shortest paths is that they
best represent the intuitive connections between any two points in a road network.

We now define the \emph{enriched graph}  $\gstar = (V,E,c)$. It consists of the
original vertices and edges and is equipped with the new cost function which implies the re-weighting of
edges. Any path finding algorithm in \gstar (e.g.\ a Dijkstra search) therefore favors
edges which are part of shortest paths between PoIs which are close according to the crowd.

When computing the cost of a whole path on \gstar, as before, we sum the respective
edge weights which differ from the original edge weights (because of the altered cost function). In
order to measure the influence of the adjusted cost values along a path $p = (e_1,\dots,e_r)$,  we
introduce the following \emph{enrichment ratio} (ER) function $er$:
 \[er(p) = \frac{1}{d(p)} \sum_{i=1}^r c(e_i)\] where $d(\cdot)$ and $c(\cdot)$ are as above. By
 normalizing with the total length of the path, we are able to compare the spatial connectivity of
 paths independent of length as well as start and target nodes. Here, a lower ratio implies higher
 closeness score values along the edges of the path.
If none of the edges of a path is part of any shortest path between PoIs,
its enrichment ratio is $1$, while the (highly unlikely) optimal enrichment ratio is $0$.

On the enriched graph \gstar \ we may now define our path finding algorithms.

\vspace{10pt}
\subsection{Path Computation on Enriched Graphs}
\label{subsec:algorithms}
\vspace{10pt}

Now that we have extracted the crowdsourced information and incorporated it into the road network,
we investigate the effect of the enrichment on the actual path computation. For this purpose, we
develop two basic algorithms which make use of the enriched network (and the relationship graph).
In Section~\ref{sec:evaluation} they are compared to the conventional shortest paths within the
original graph, as obtained with Dijkstra's algorithm, which we denote by \gd.

Note that all paths in this paper are computed by Dijkstra's algorithm 
because our main focus is not the routing itself but the incorporation of textual information into existing road
networks. If desired, speed-up techniques, such as preprocessing steps and/or other search
algorithms, could easily be employed.

We refer to the first approach we want to present as \gstard. Given start and target nodes, it
executes a Dijkstra search in the enriched road network graph \gstar \ w.r.t. the adjusted cost
function.

Our second approach, referred to as \hstard, uses the enriched road network graph \gstar \ as well
as the relationship graph \hstar. Given start and target nodes within \gstar, entry and exit nodes
within \hstar \ are determined. Subsequently, we route within \hstar, i.e. from PoI to PoI,
again using Dijkstra's algorithm.

The pseudo-code for the second approach is given in Algorithm~\ref{alg:hpg}. All three approaches
-- \gd, \gstard \ and \hstard -- return paths connecting start and target. But while \gd~ computes
the shortest path in the original graph \g, \gstard \ computes the shortest path in the enriched
graph w.r.t. the adjusted cost function $c$.
By construction of $c$, it favors edges which are part of shortest paths between close
PoIs. \hstard \ in contrast, does not only favor these edges, but is restricted to them.
Having found entry and exit nodes within \hstar, \hstard \ hops from PoI to PoI in
direction of the target. Hence, \gd, \gstard, \hstard, in that order, represent an increasing binding to the
extracted relations. \gd \ is not bound to the relations at all, while \gstard \ (by the adjusted
cost function) favors ``relation-edges'', and \hstard \ is strictly bound to the relations and the
graph formed by them. The difference in output is also visualized in Figure~\ref{fig:intuition} where
the continuous line represents the shortest path from start to target provided by \gd, while the dot dashed and dotted lines represent alternative paths
computed by the \gstard and \hstard, respectively. 

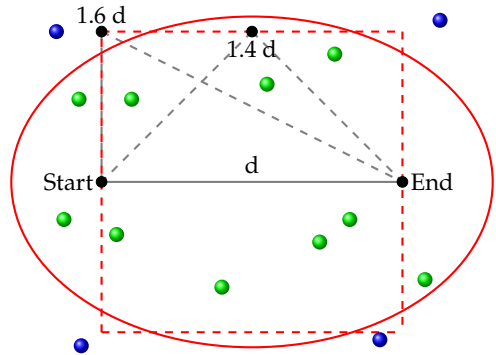
\begin{figure}[!h]
\begin{center}
 \begin{tikzpicture}
\draw[gray, thick,dashed] (5,2.0) -- (3,0);
\draw[gray, thick,dashed] (5,2.0) -- (7,0);
\draw[gray, thick,dashed] (3,2.0) -- (7,0);
\draw[gray, thick,dashed] (3,2.0) -- (3,0);

\draw[red, thick,dashed] (3,-2) -- (3,2);
\draw[red, thick,dashed] (7,2) -- (7,-2);
\draw[red, thick,dashed] (3,2) -- (7,2);
\draw[red, thick,dashed] (3,-2) -- (7,-2);

\draw[gray, thick] (3,0) -- (7,0);
\filldraw[black] (3,0) circle (2pt) node[anchor=east] {Start};
\filldraw[black] (7,0) circle (2pt) node[anchor=west] {End};
\filldraw[black] (5,2.0) circle (2pt) node[anchor=north] {1.4 d};
\filldraw[black] (3,2.0) circle (2pt) node[anchor=south]{1.6 d};

\draw (5, 0) node[anchor=south] {d};
\draw[red, thick] (5,0) ellipse (3.2cm and 2.2cm);

\shade [ball color = green] (2.7,1.1) circle (0.1);
\shade [ball color = green] (3.2,-0.7) circle (0.1);
\shade [ball color = green] (6.3,-0.5) circle (0.1);
\shade [ball color = green] (5.2,1.3) circle (0.1);
\shade [ball color = green] (4.6,-1.4) circle (0.1);
\shade [ball color = green] (2.5,-0.5) circle (0.1);
\shade [ball color = green] (6.1,1.7) circle (0.1);
\shade [ball color = green] (7.3,-1.3) circle (0.1);
\shade [ball color = green] (3.4, 1.1) circle (0.1);
\shade [ball color = green] (5.9,-0.8) circle (0.1);

\shade [ball color = blue] (2.4,2.0) circle (0.1);
\shade [ball color = blue] (2.73,-2.18) circle (0.1);
\shade [ball color = blue] (7.5,2.15) circle (0.1);
\shade [ball color = blue] (6.7,-2.1) circle (0.1);

\end{tikzpicture}
\caption{Restriction of relationship graph \hstar \ to a subgraph $H^{**}$, in order to avoid
implausible detours. The green dots represent PoIs, i.e.\ nodes of \hstar \ which are also in
$H^{**}$, the blue ones are left out.}
\label{fig:validpois}
\end{center}
\end{figure}

Here, let us formalize \hstard. Given start and target node in \gstar, it
first determines the so-called entry and exit nodes to and from \hstar. However, to exclude PoIs
which would imply a significant detour, we restrict the set of valid PoIs, i.e. we restrict the
search to a subgraph of \hstar, denoted as $H^{**}$. Figure \ref{fig:validpois} illustrates our
computationally inexpensive implementation of a query ellipse that allows for some deviation in the
middle of the path as well as for initial and final detours.

%
%
%
%
%
%

\IncMargin{1em}
\begin{algorithm}[!tb]
\DontPrintSemicolon
 \KwIn{Enriched Graph \gstar, Relationaship Graph \hstar, start $s$, target $t$}
 \KwOut{Path $p$ between $s$ and $t$}
 \BlankLine
 \Begin{

$H^{**} \leftarrow$ subgraph of \hstar \ in bounding ellipse

$p \leftarrow$ empty path

\BlankLine

$P_\text{entry} \leftarrow$ select PoI $P \in H^{**}$ closest to $s$

$P_\text{exit} \leftarrow$ select PoI $P \in H^{**}$ closest to $t$

$h \leftarrow \text{Dijkstra}(H^{**}, P_\text{entry}, P_\text{exit})$ \;

\BlankLine

$\text{predecessor} \leftarrow s$

\ForEach{PoI $P$ on path $h$}{
	$v \leftarrow$ select node $v \in \gstar$ representing $P$
	
	$p.\proc{Append}(\text{Dijkstra}(\gstar, \text{predecessor}, v))$
	
	$\text{predecessor} \leftarrow v$
}

$p.\proc{Append}(\text{Dijkstra}(\gstar, last, t))$ \;

\BlankLine

\Return{$p$} \;

\BlankLine

}%
\caption{\hstard}
\label{alg:hpg}
\end{algorithm}

\vspace{10pt}

After selecting the valid set of PoIs
(Step 2), entry and exit nodes to and from \hstar \ are determined, i.e. the closest PoIs to start
and target node, respectively (Steps 4 and 5). Entry and exist nodes connect the road network \gstar
\ to the relationship  graph \hstar. Subsequently, the shortest path in $H^{**}$ from entry to exit node is
computed  using Dijkstra's algorithm w.r.t. the Euclidean distance (Step 5). Note that a shortest
path  within \hstar \ is a sequence of PoIs. We therefore map this sequence
onto  \gstar \ by computing the shortest paths between the consecutive pairs of PoIs in \gstar \
w.r.t. the adjusted cost function (Step 8). Also, we compute the shortest paths in \gstar \ from
start to entry node and exit to target node. Concatenating these paths (start to entry, PoI to
PoI, exit to target), we return a full path.

\section{Experimental Evaluation}
\label{sec:evaluation}

Implementing the algorithms presented above, we want to investigate the effect and impact of the
network enrichment. We compare the results of the conventional Dijkstra search, \gd, to the results
of \gstard, which uses the enriched graph \gstar, and the results of \hstard, which mainly relies on
the relationship graph \hstar, all evaluated on real world datasets. Besides comparing the computed path w.r.t.\
their enrichment ratio (ER) and length (as presented in Section~\ref{subsec:enrich}), we introduce a
measure of popularity based on Flickr data, which is explained in the following section.
All the text processing parts were implemented in Python while modeling parts were implemented in Matlab.
Network enrichment and path computation tasks were conducted using the Java-based MARiO Framework
\cite{GraKriRenSch11} on an Intel(R) Core(TM) i7-3770 CPU at 3.40GHz and 32 GB RAM running Linux (64 bit).

\vspace{10pt}
\subsection{Enrichment Ratio, Distance and Popularity Evaluation}

\begin{table*}[!t]
\centering
\caption{Statistics for the relationship graphs, Flickr datsets and road networks of Paris, New York
and London, respectively.}
\ra{1.1}
\begin{tabular}{l || c c c c || c c c c || c c c c}
\toprule
& \multicolumn{4}{c||}{Relationship Graph ($\hstar$)} & \multicolumn{4}{c||}{Flickr} &
\multicolumn{4}{c}{Road Network ($\g$)} \\
\toprule
Dataset && \# PoI Pairs &&  \# Relations && \# Photos && \# Max Photos per Vertex && \# Vertices &&  \# Edges \\

\cmidrule{1-1} \cmidrule{3-3} \cmidrule{5-5} \cmidrule{7-7} \cmidrule{9-9} \cmidrule{11-11} \cmidrule{13-13}
Paris       &&  400     &&  2000    && 400$K$    &&  100   &&    550$K$     && 300$K$\\
New York    &&  300     &&  1500    &&  90$K$        &&  200   &&    220$K$    && 120$K$\\
London      &&  150     &&  1000    && 70$K$     &&  150   &&    220$K$    && 110$K$\\
\bottomrule
\end{tabular}\label{table:stats}
\end{table*}

Our experimental setup is based on three regions, Paris, London and New York. These regions
have comparatively high density of spatial relations, Flickr photo data, and OSM data, which
accounts for an exact representation of the road networks. As mentioned before, we compare the
output of \gd, \gstard, and \hstard \ w.r.t.\ to the paths they return, more precisely, w.r.t.\
ER and length of these paths. Since ER is a measure introduced in this paper,
we use Flickr data as an independent ground truth. We are aware that to
cognitive aspects (like the importance of sights or the value of landmarks) there is no absolute
truth. However, in order to be able to draw comparisons, we presume that if the dataset is large
enough, the bias can be neglected. We use a geotagged Flickr photo dataset, provided by the
authors in \cite{Flickr}, to assign a number of photos to each vertex of the underlying road
network. The number of Flickr photos assigned to each vertex is referred to
\emph{popularity}. In our settings, every photo which is within the $20$-meter radius of a
vertex, contributes to the popularity of that vertex. The popularity of a path is
computed by summation of all popularity values along this path.

The sizes of the relationship graph, road network and Flickr photo data for all three cities
are shown in Table~\ref{table:stats}. Regarding the relationship graphs, we provide the number of
unique PoI pairs extracted from the travel blog corpus and the number of spatial (closeness) relations
extracted between them, as was analyzed in Section~\ref{sec:contribution}. Regarding Flickr data,
we provide the total number of geotagged photos in each city and the maximum number of
photos assigned to one vertex of the road network. Finally, regarding the road network,
we provide the total number of edges and vertices. Note that although the datasets differ in terms
of density (w.r.t.\ to relations and Flickr photos), our algorithms provide similar results.

Continuing, we present two experimental settings: In Setting $(i)$ we examine the influence of
different scalings of the closeness score $\wij$ in terms of enrichment ratio, path length increase (distance) and
popularity. Setting $(ii)$ investigates the deviation of shortest paths in
the original road network with shortest paths in the enriched graph. In both settings we present
the ER performance of the two algorithms separately from their performance in terms of distance
and popularity as ER is a measure that mainly proves that our network enrichment approach
works properly, i.e., ER should increase with the increase of the influence of $\wij$ on the network
and the increase of the path length.

\begin{figure}[!t]

%
%
%
%

\subfigure[]{\includegraphics[width=0.33\textwidth]{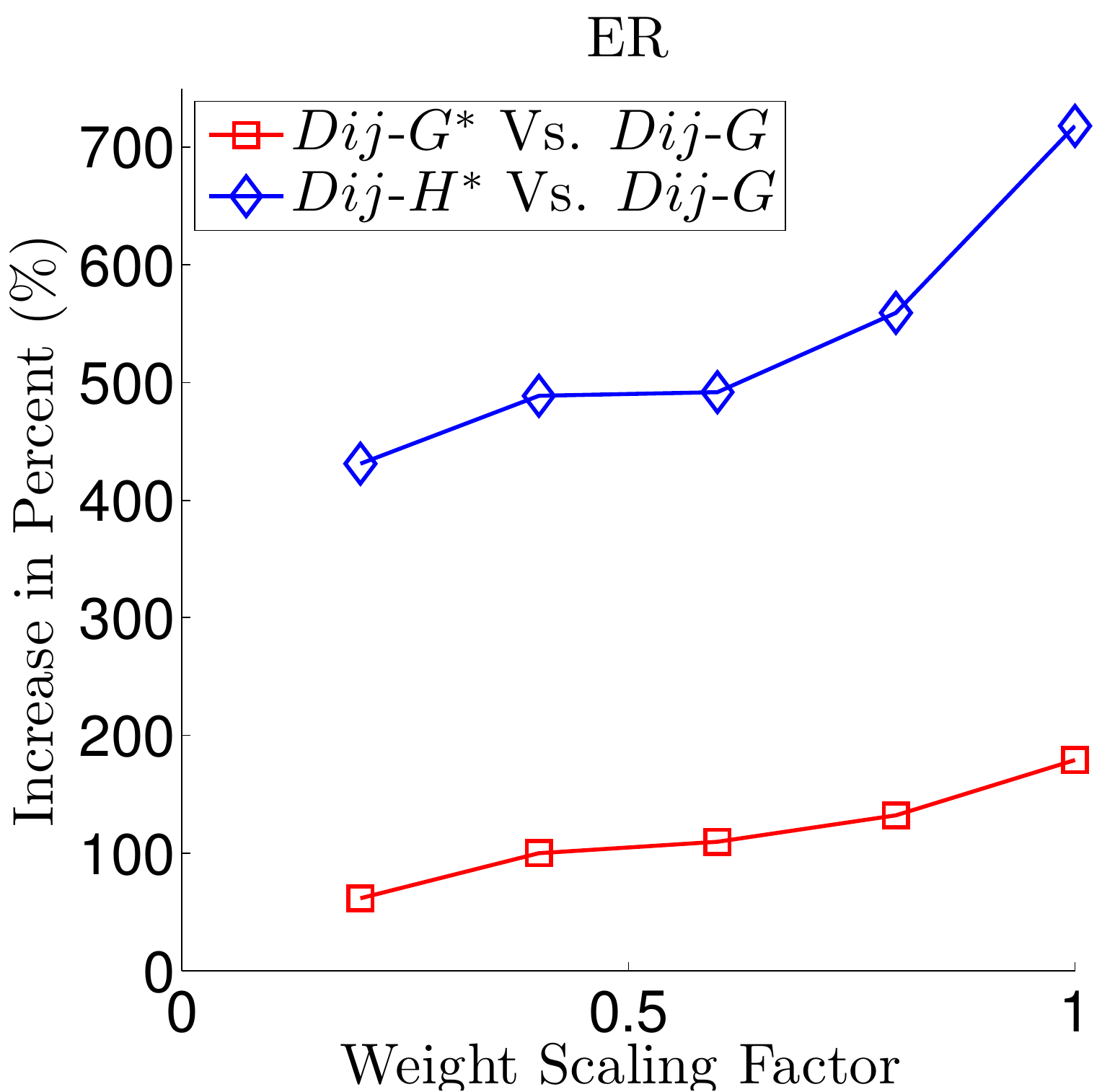}\label{subfig:ParisScsVsWeight}}
\subfigure[]{\includegraphics[width=0.33\textwidth]{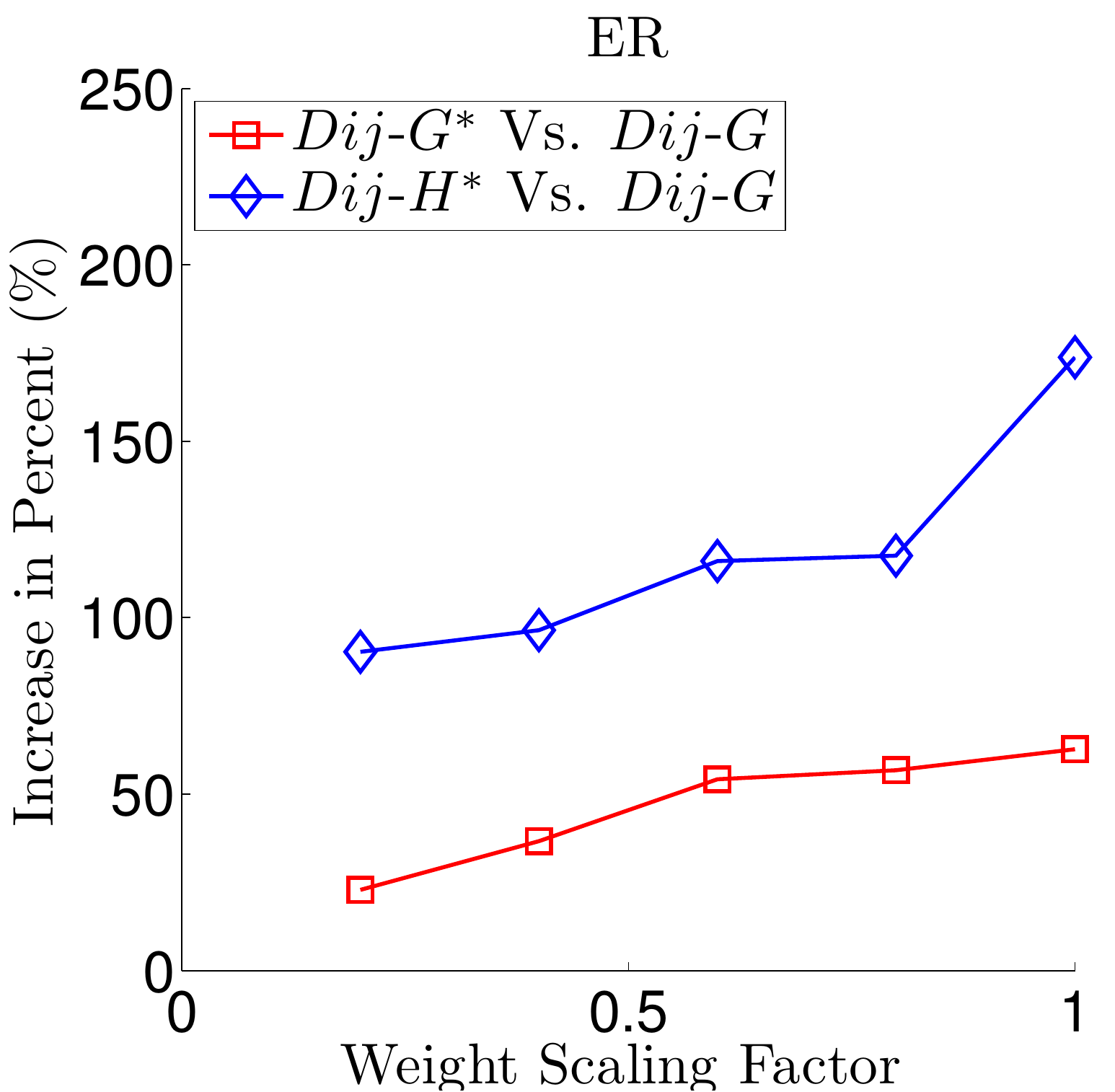}\label{subfig:NYScsVsWeight}}
\subfigure[]{\includegraphics[width=0.33\textwidth]{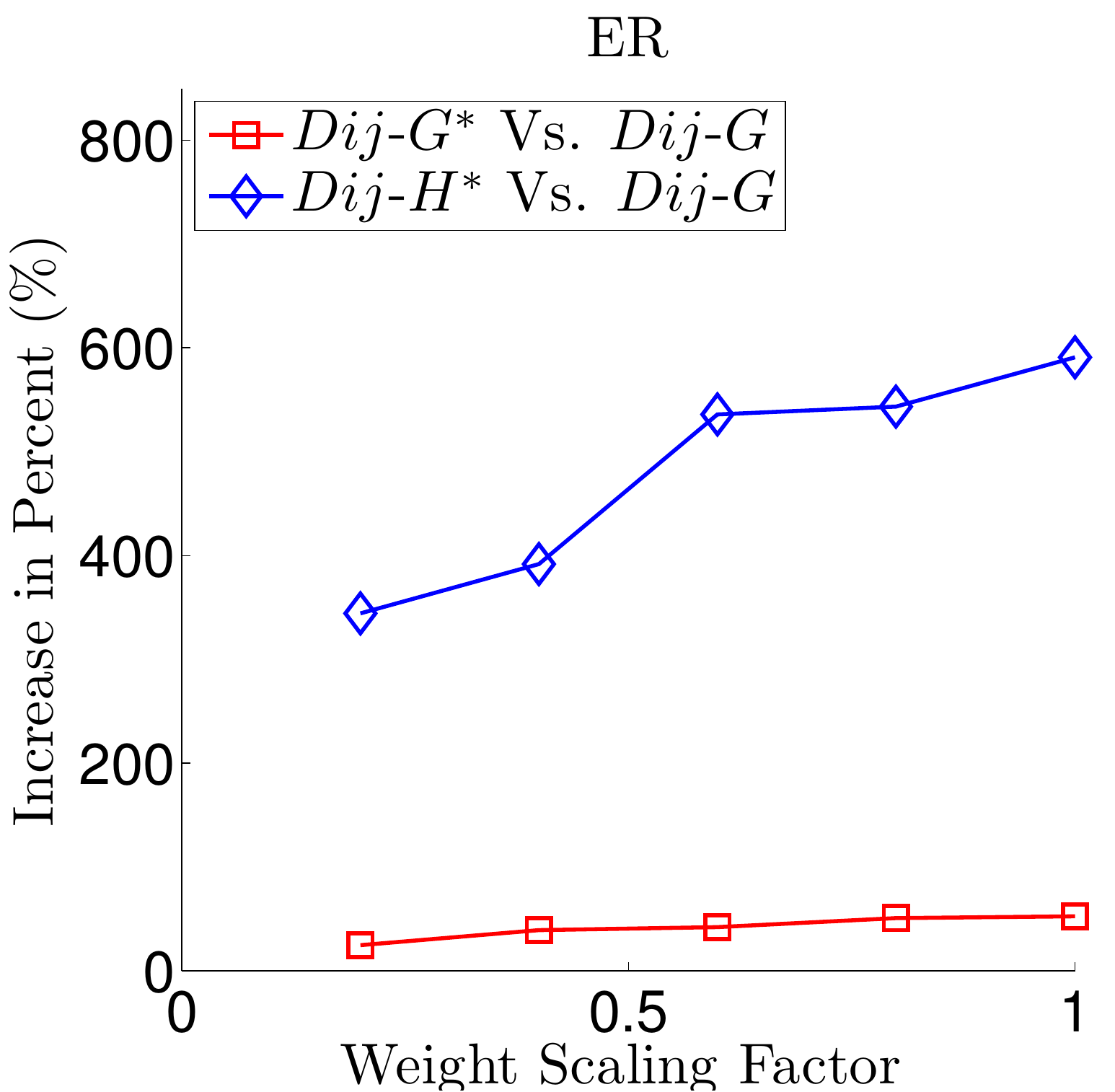}\label{subfig:LondonScsVsWeight}}\\

\subfigure[]{\includegraphics[width=0.33\textwidth]{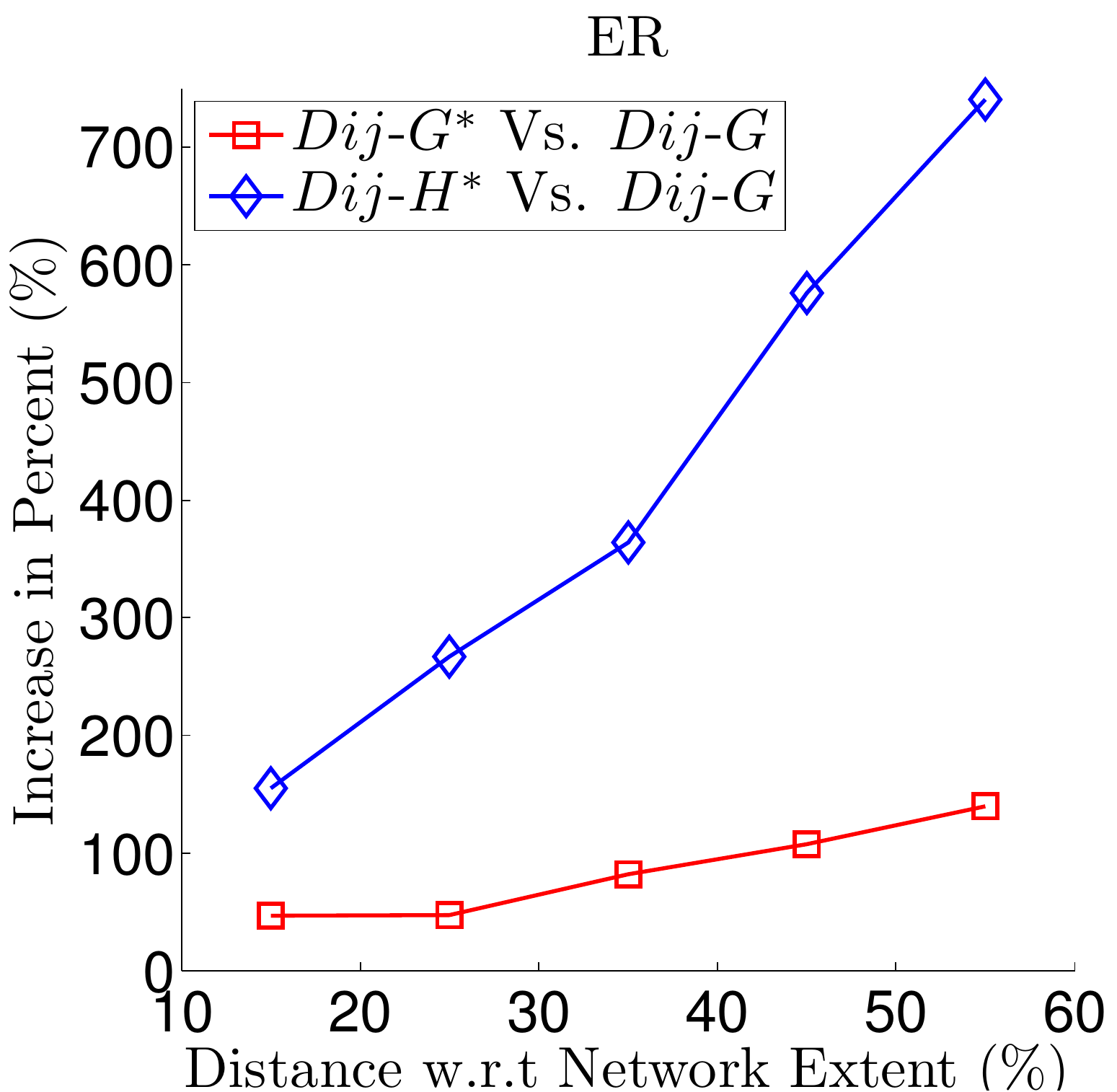}\label{subfig:ParisScsVsDistExt}}
\subfigure[]{\includegraphics[width=0.33\textwidth]{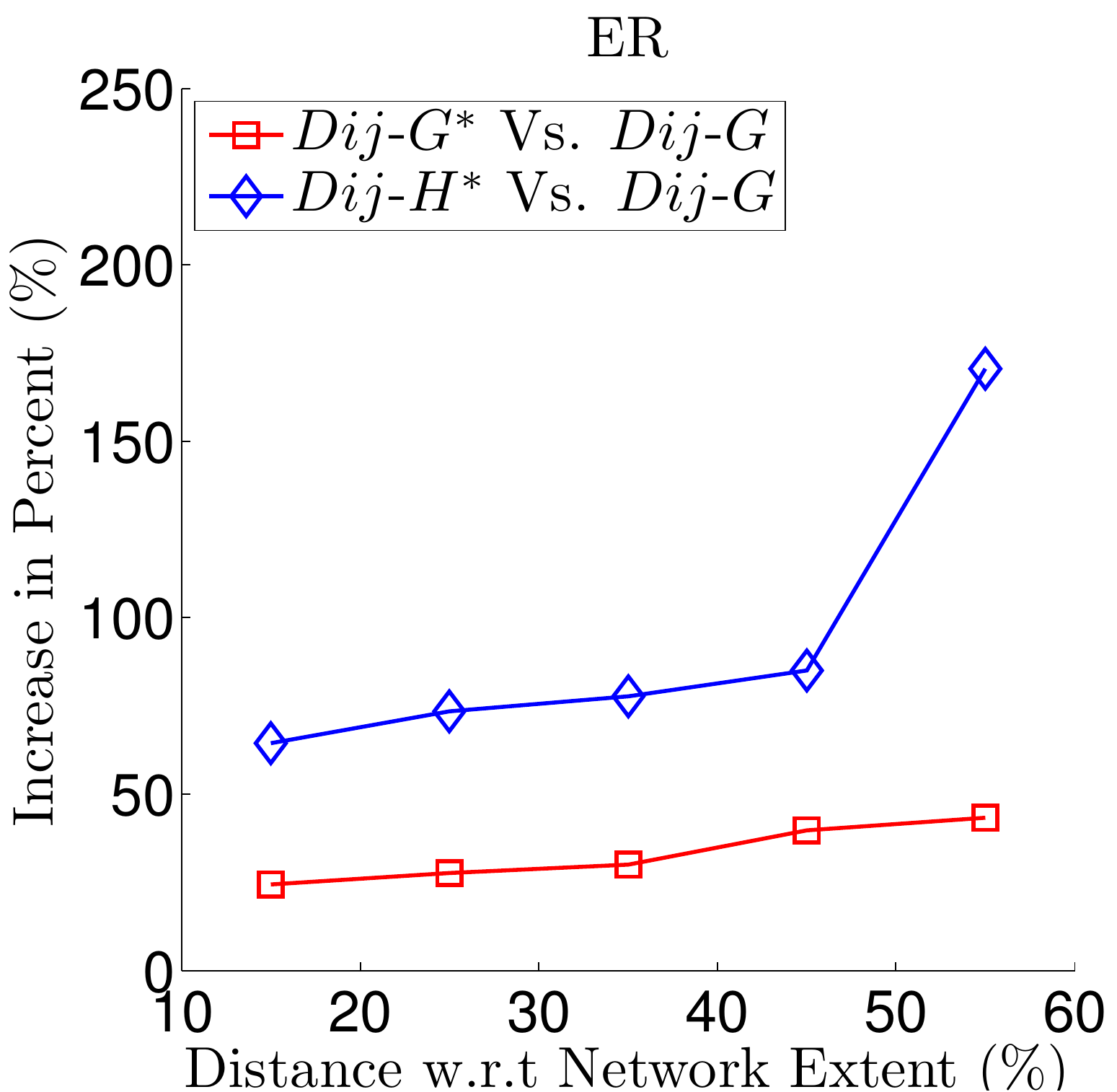}\label{subfig:NYScsVsDistExt}}
\subfigure[]{\includegraphics[width=0.33\textwidth]{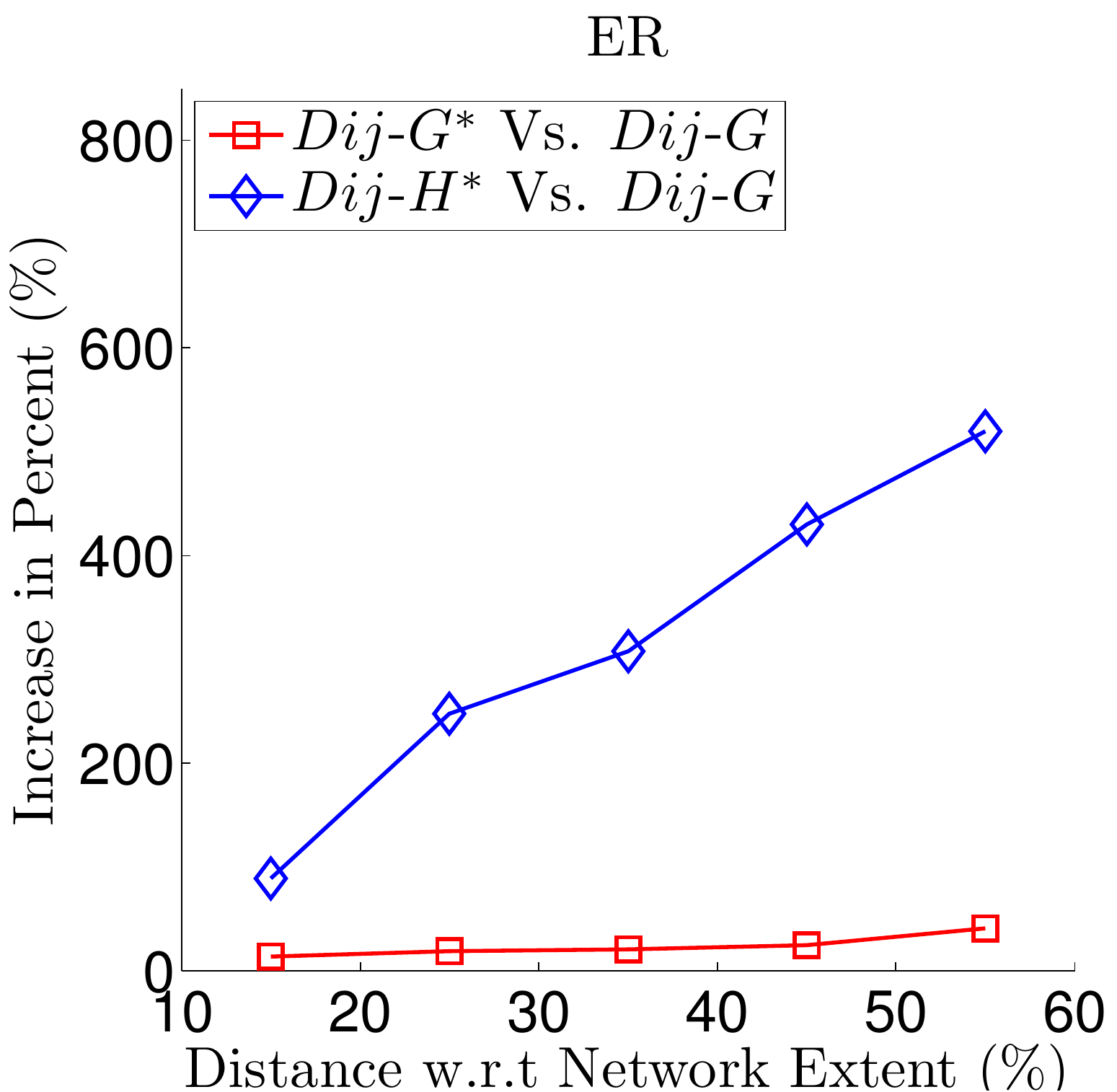}\label{subfig:LondonScsVsDistExt}}

\caption{(a), (b) show ER increase for algorithms \gstard \ and \hstard for Paris dataset for Settings $i$ and $ii$ respectively.
(c), (d) show ER increase for algorithms \gstard \ and \hstard for New York dataset for Settings $i$ and $ii$ respectively.
(e), (f) show ER increase for algorithms \gstard \ and \hstard for London dataset for Settings $i$ and $ii$ respectively.}
\label{fig:allER}
\end{figure}
\begin{figure}[!t]
\subfigure[]{\includegraphics[width=0.245\textwidth]{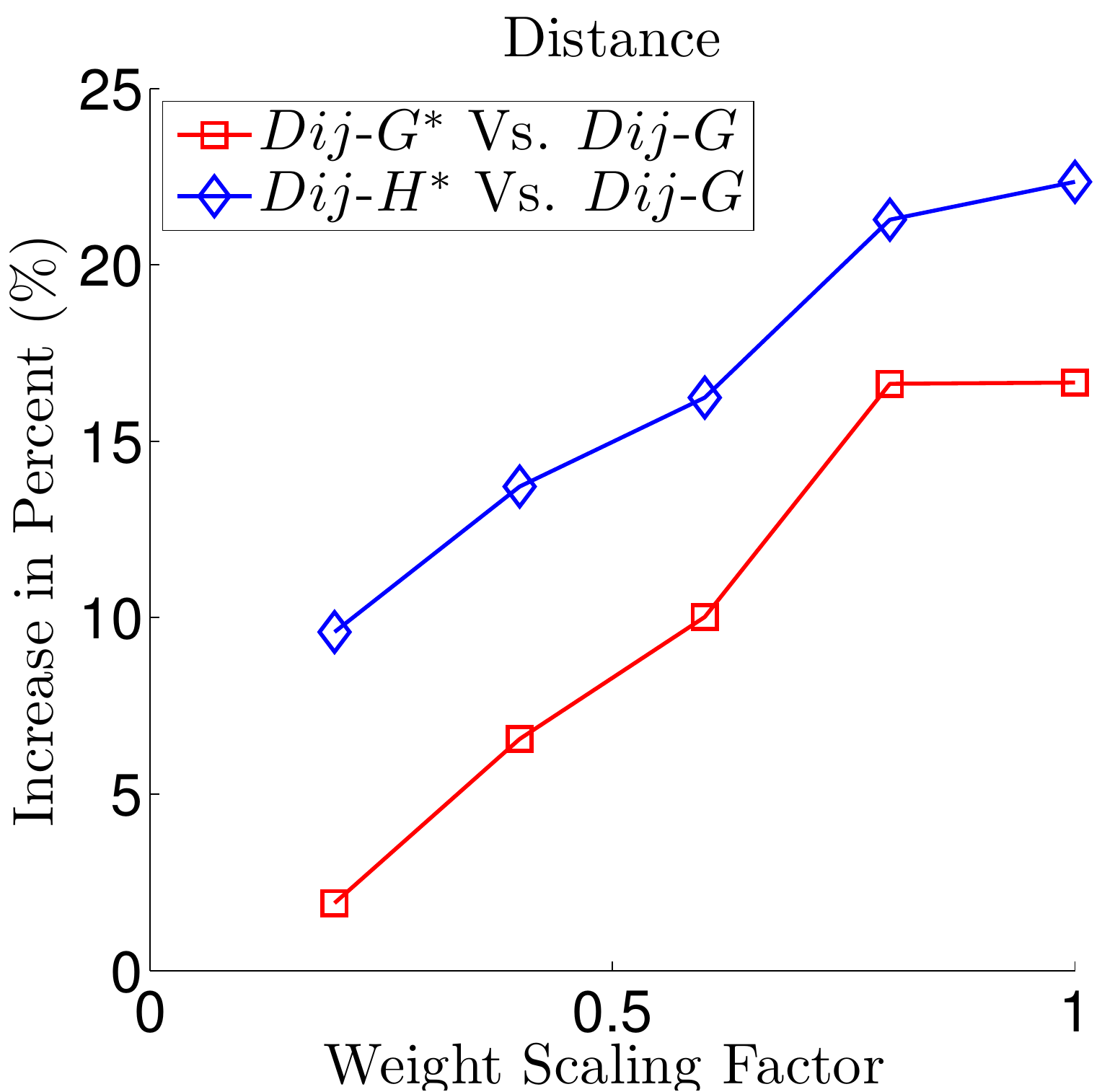}\label{subfig:ParisDistVsWeight}}
\subfigure[]{\includegraphics[width=0.245\textwidth]{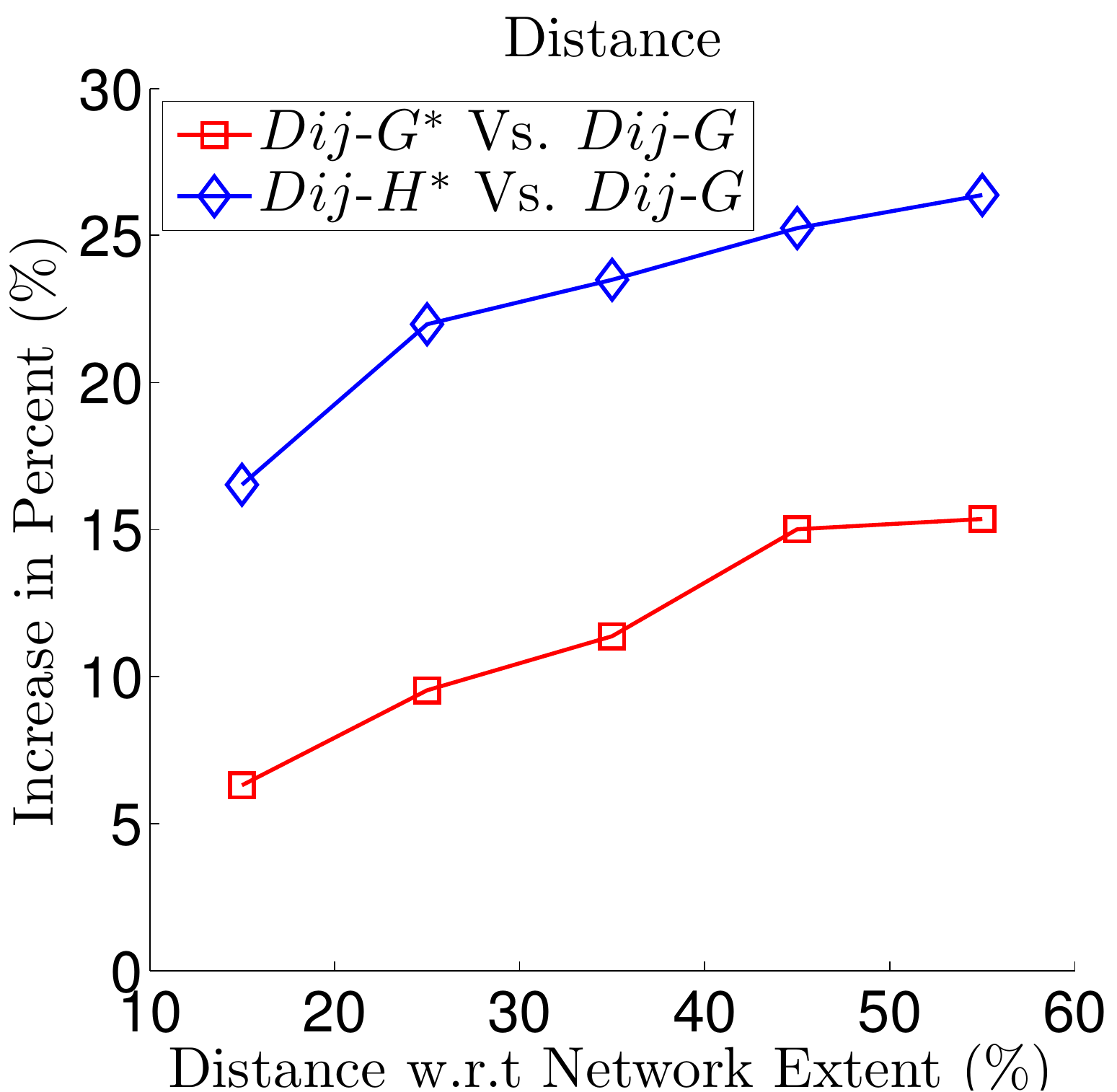}\label{subfig:ParisDistVsDistExt}}
\subfigure[]{\includegraphics[width=0.245\textwidth]{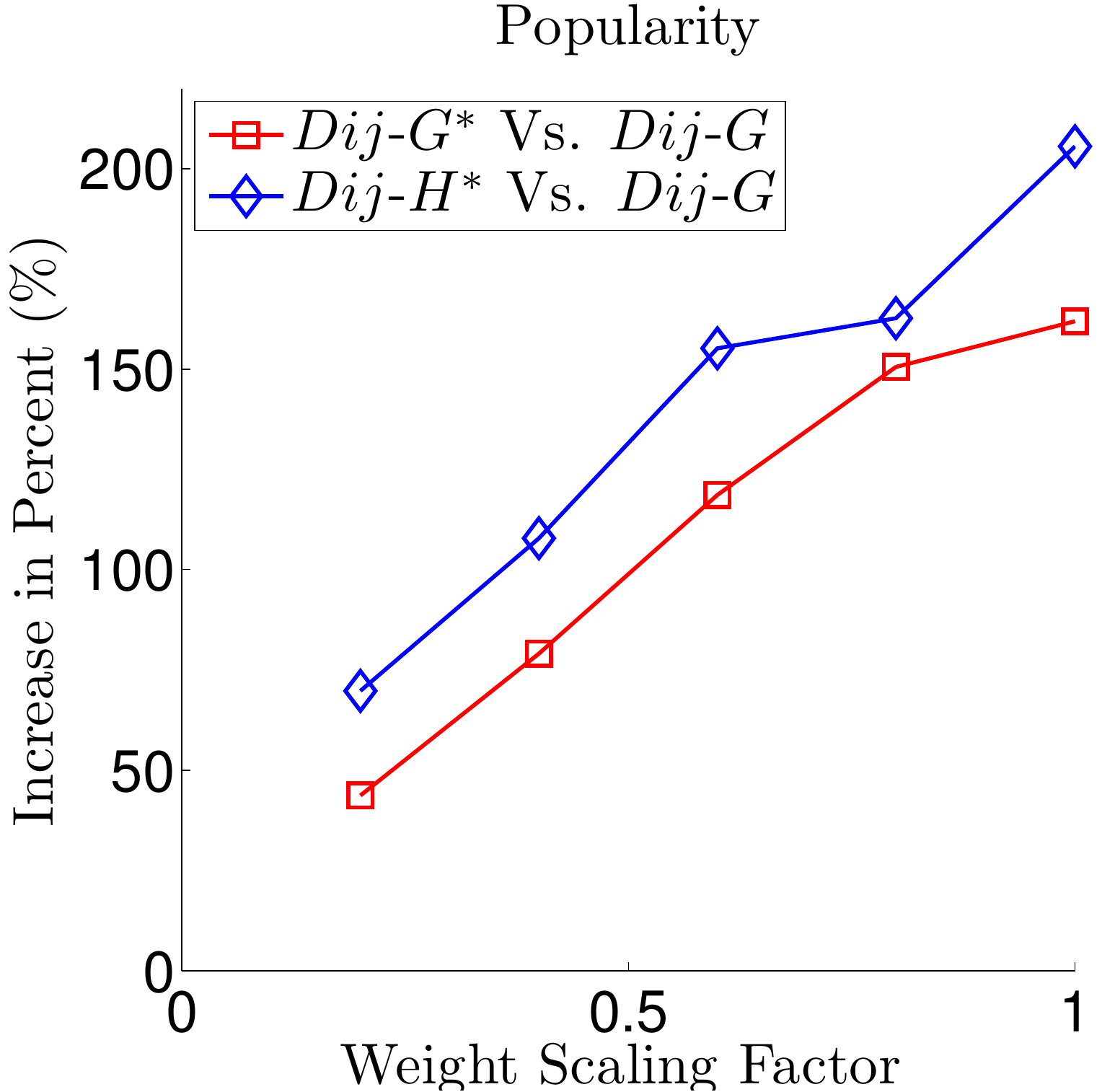}\label{subfig:ParisFlickrVsWeight}}
\subfigure[]{\includegraphics[width=0.245\textwidth]{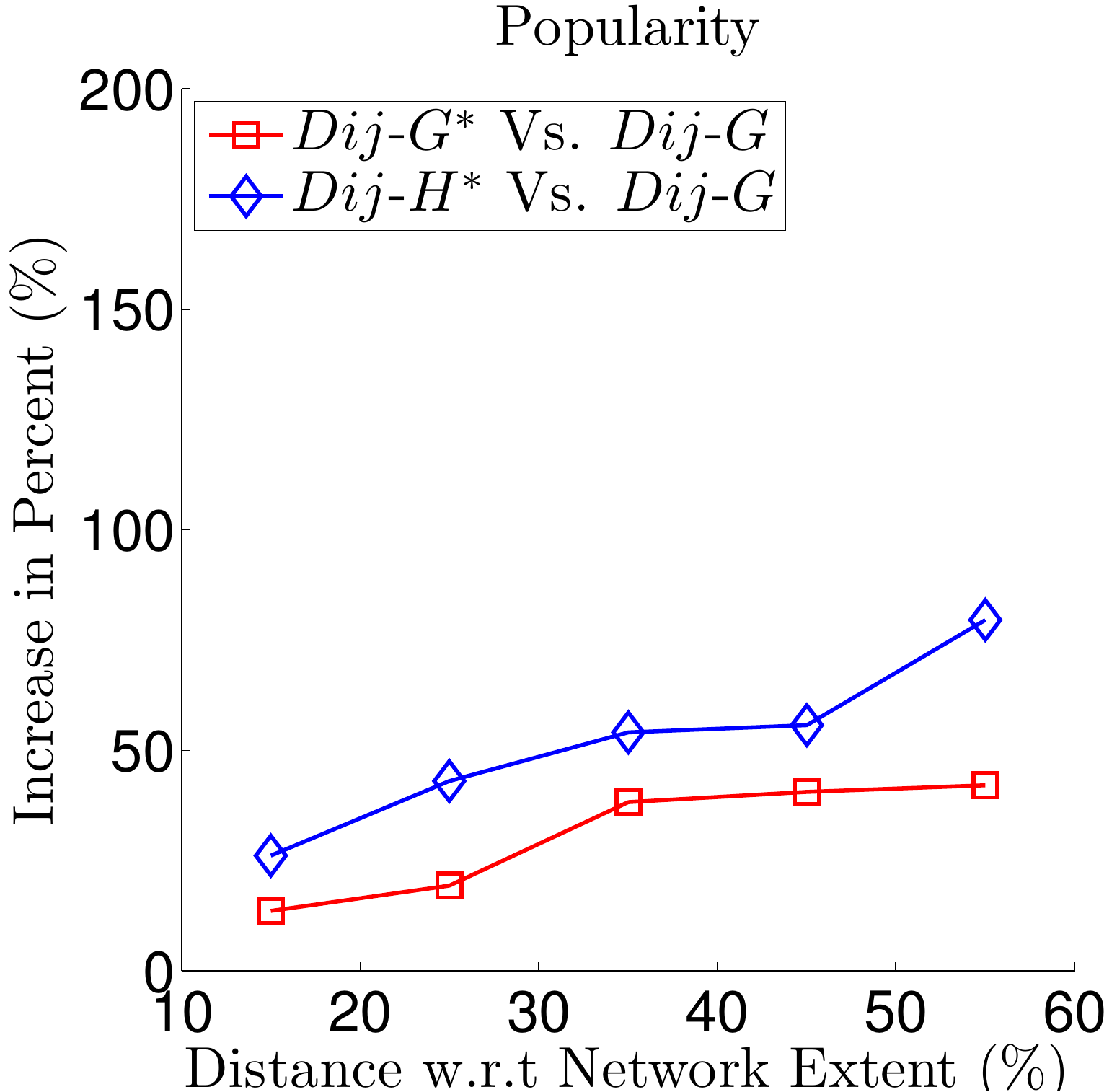}\label{subfig:ParisFlickrVsDistExt}}
\caption{(a), (c) show Distance and Flickr popularity increase for algorithms \gstard \ and \hstard \ for Paris dataset for experimental Setting $i$.
(b), (d) show Distance and Flickr popularity increase for algorithms \gstard \ and \hstard \ for Paris dataset for experimental Setting $ii$.}
\label{fig:paris}
\end{figure}
\begin{figure}[!ht]
\subfigure[]{\includegraphics[width=0.245\textwidth]{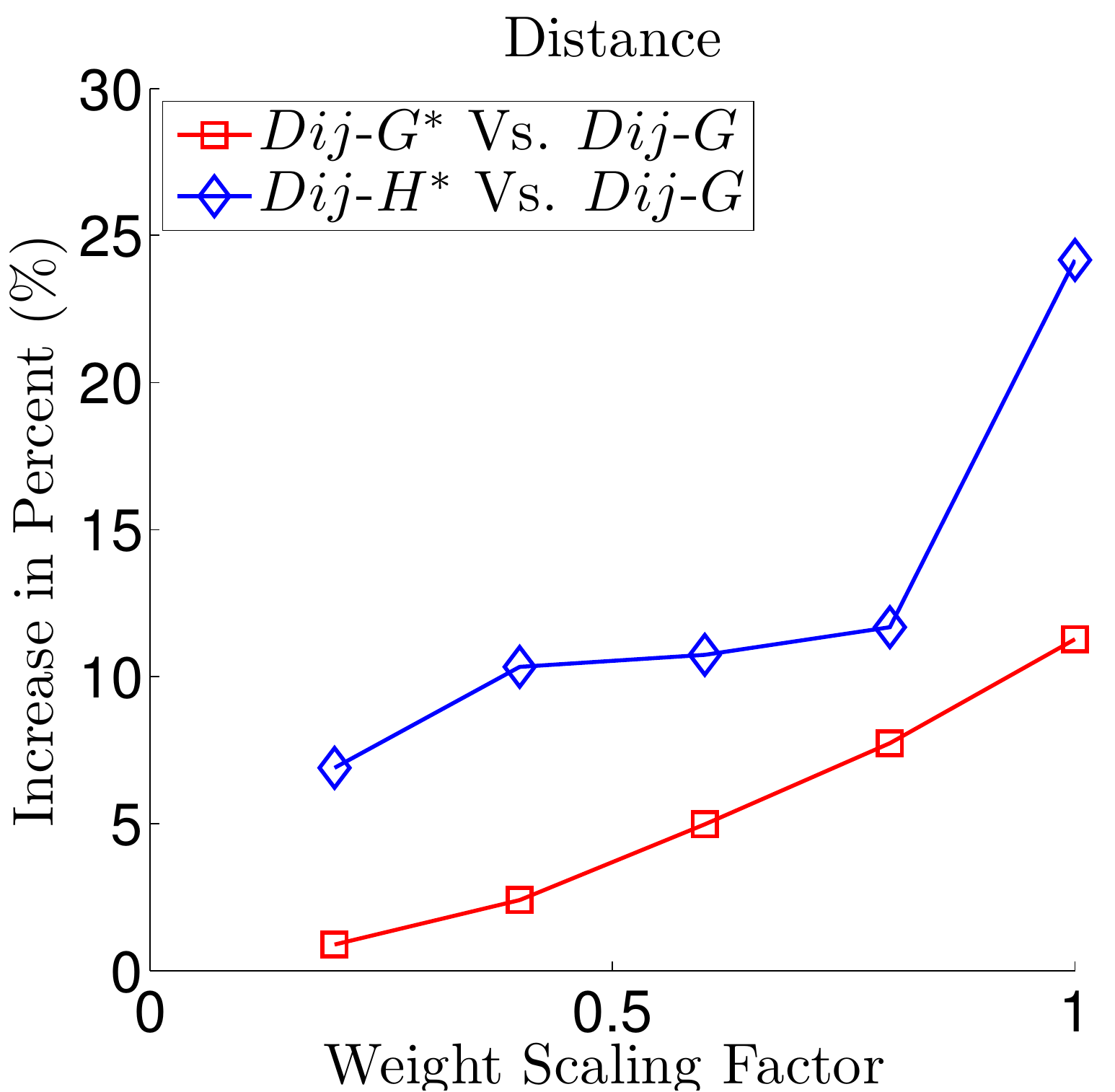}\label{subfig:NYDistVsWeight}}
\subfigure[]{\includegraphics[width=0.245\textwidth]{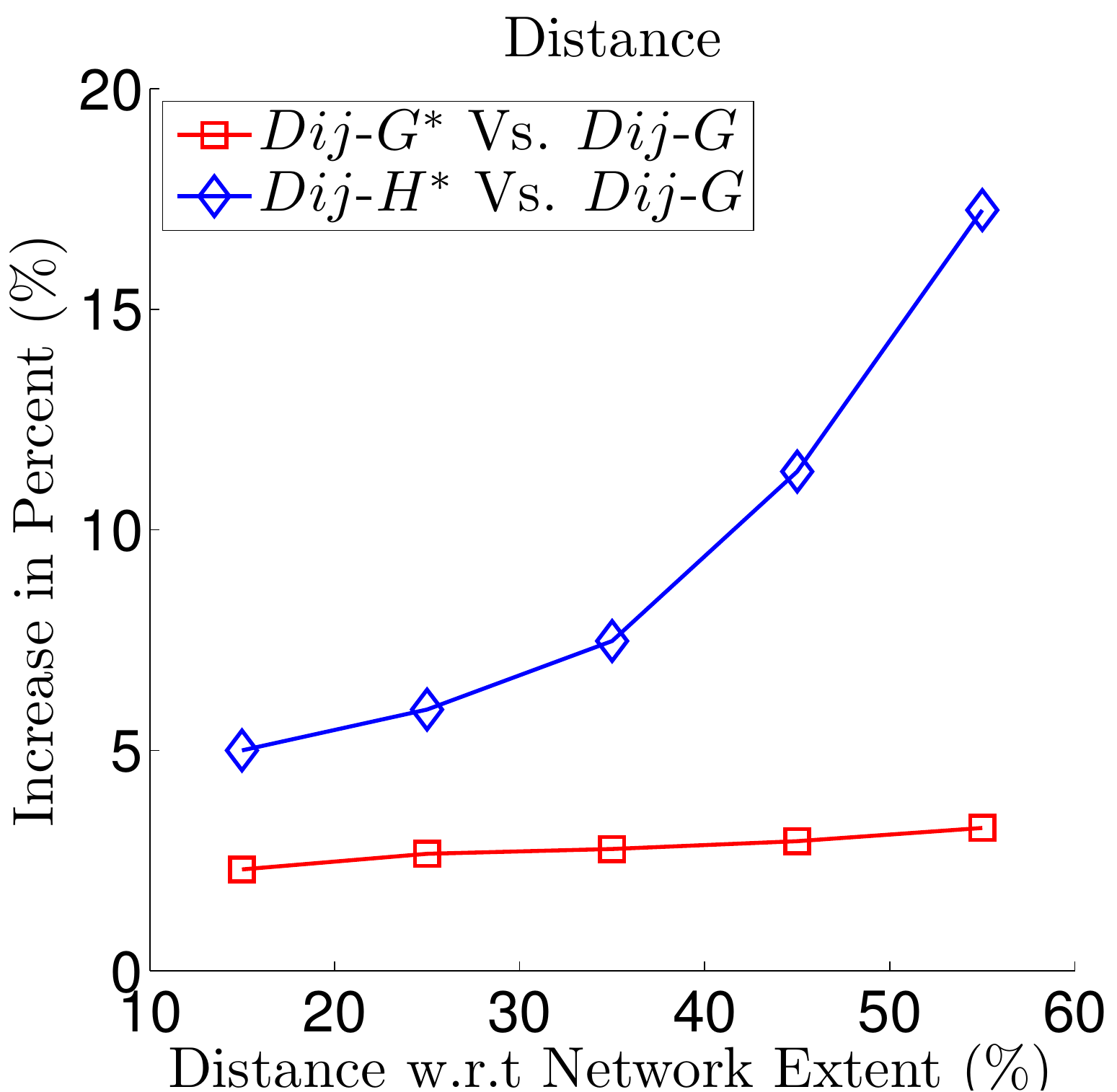}\label{subfig:NYDistVsDistExt}}
\subfigure[]{\includegraphics[width=0.245\textwidth]{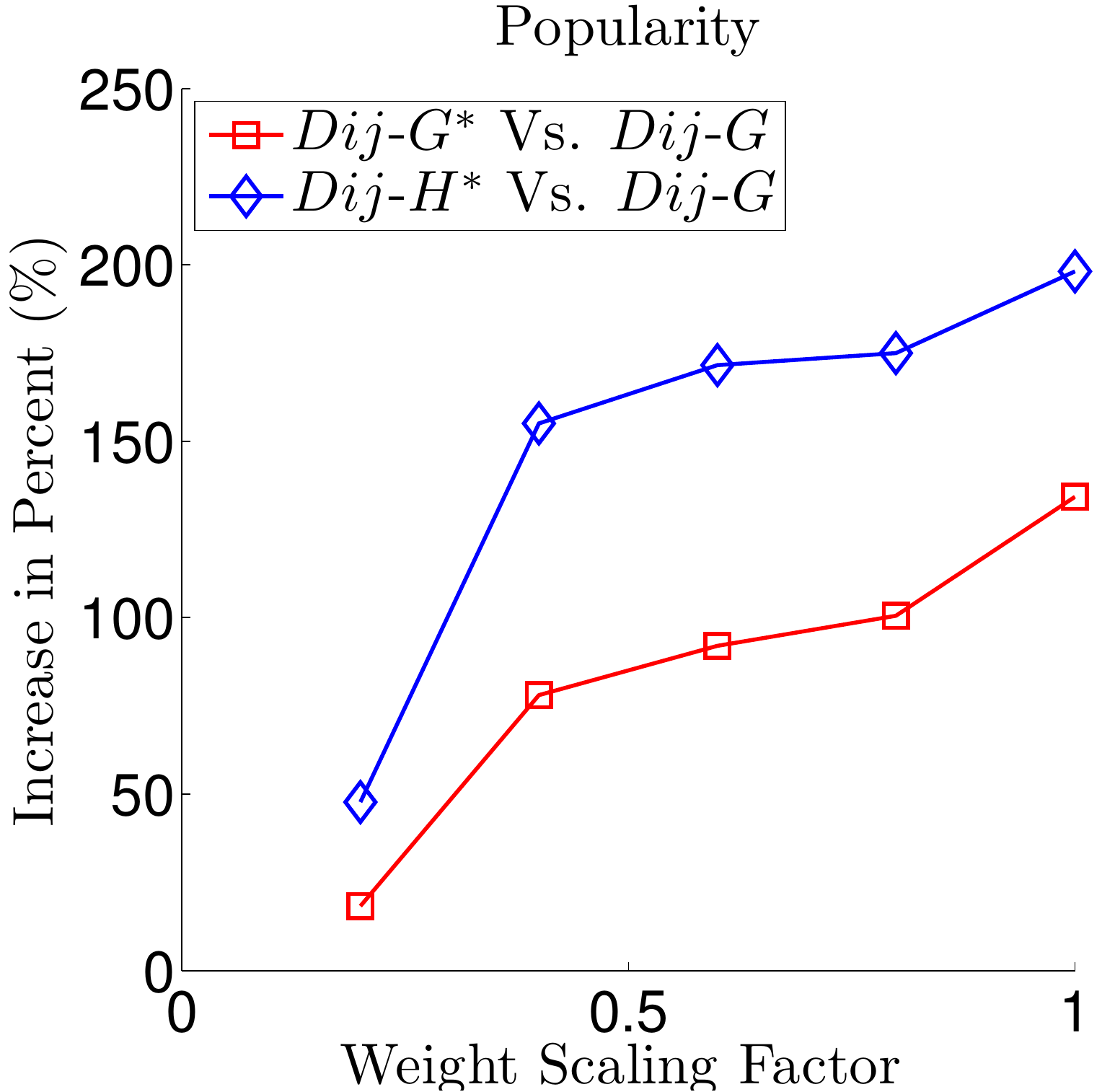}\label{subfig:NYFlickrVsWeight}}
\subfigure[]{\includegraphics[width=0.245\textwidth]{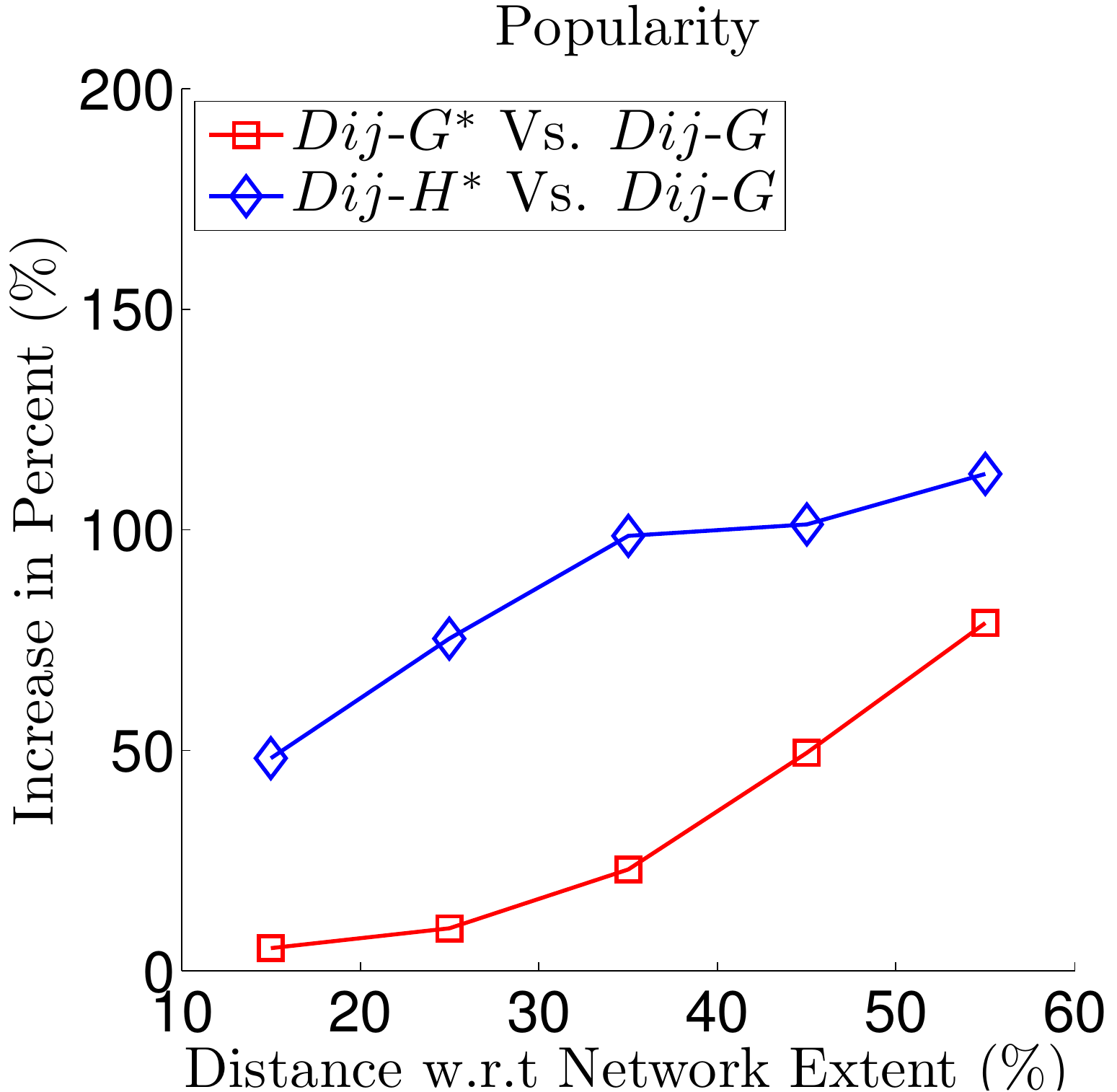}\label{subfig:NYFlickrVsDistExt}}
\caption{(a), (c) show Distance and Flickr popularity increase for algorithms \gstard \ and \hstard \ for New York dataset for experimental Setting $i$.
(b), (d) show Distance and Flickr popularity increase for algorithms \gstard \ and \hstard \ for New York dataset for experimental Setting $ii$.}
\label{fig:ny}
\end{figure}
\begin{figure}[!ht]
\subfigure[]{\includegraphics[width=0.245\textwidth]{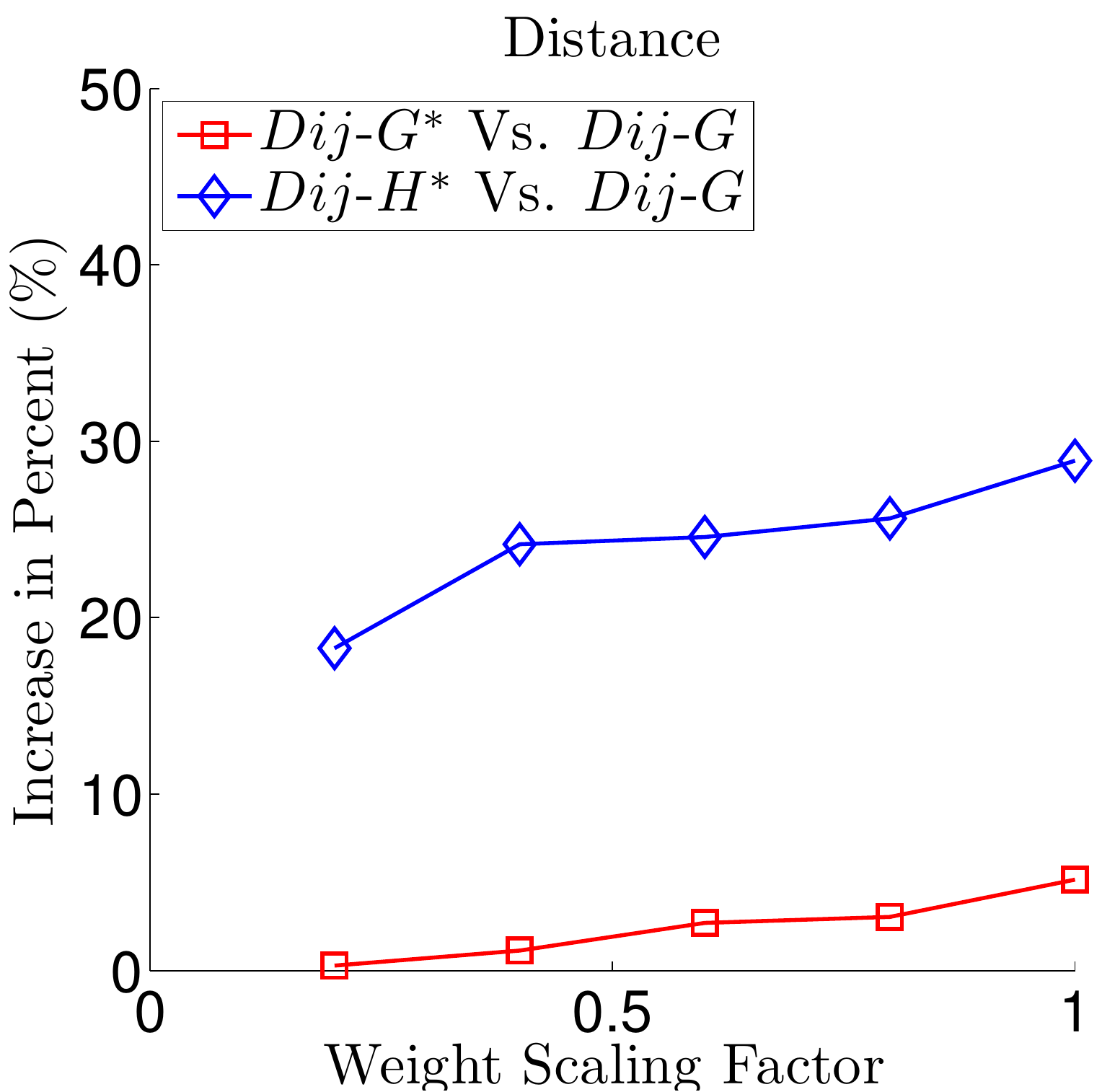}\label{subfig:LondonDistVsWeight}}
\subfigure[]{\includegraphics[width=0.245\textwidth]{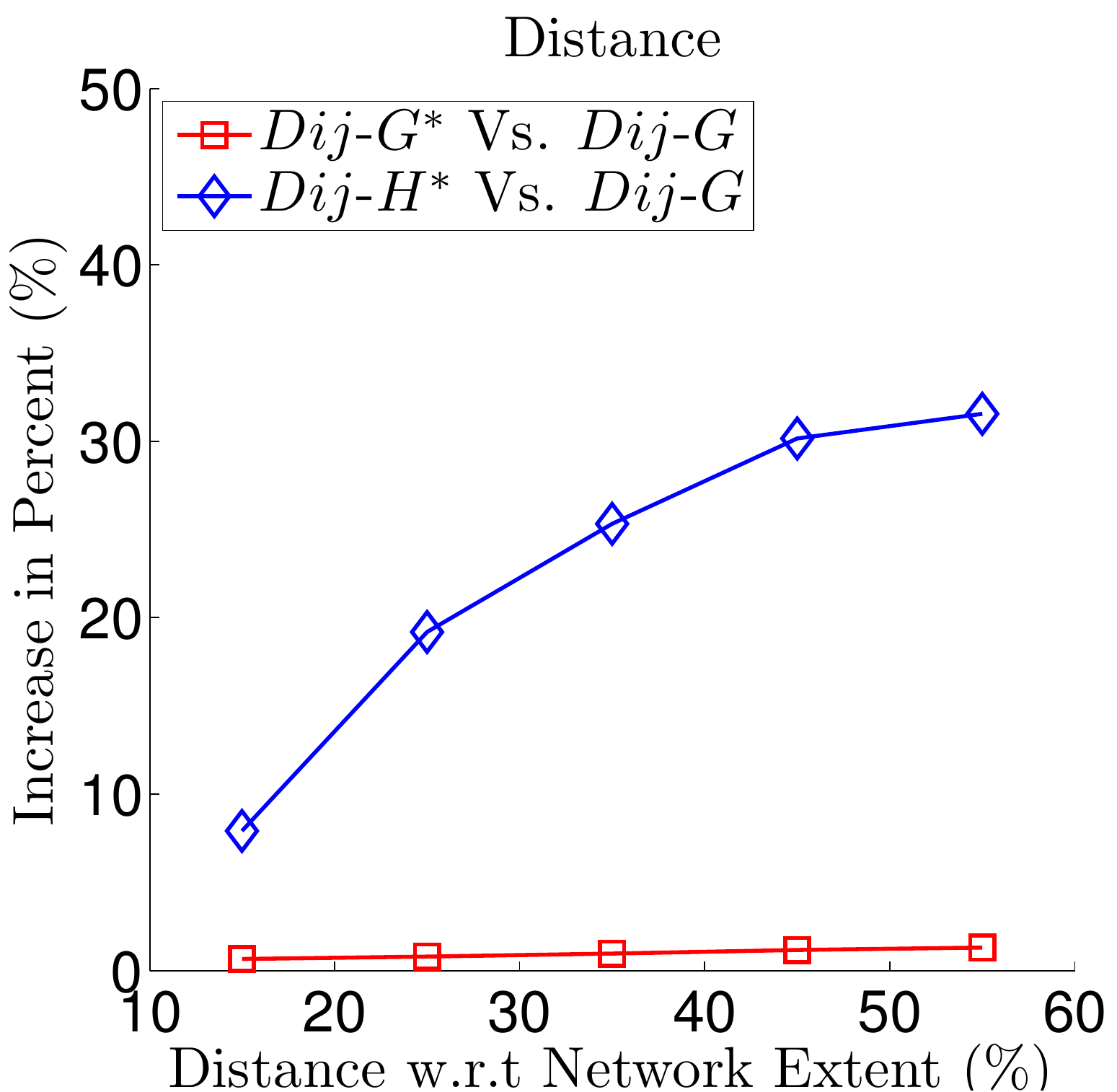}\label{subfig:LondonDistVsDistExt}}
\subfigure[]{\includegraphics[width=0.245\textwidth]{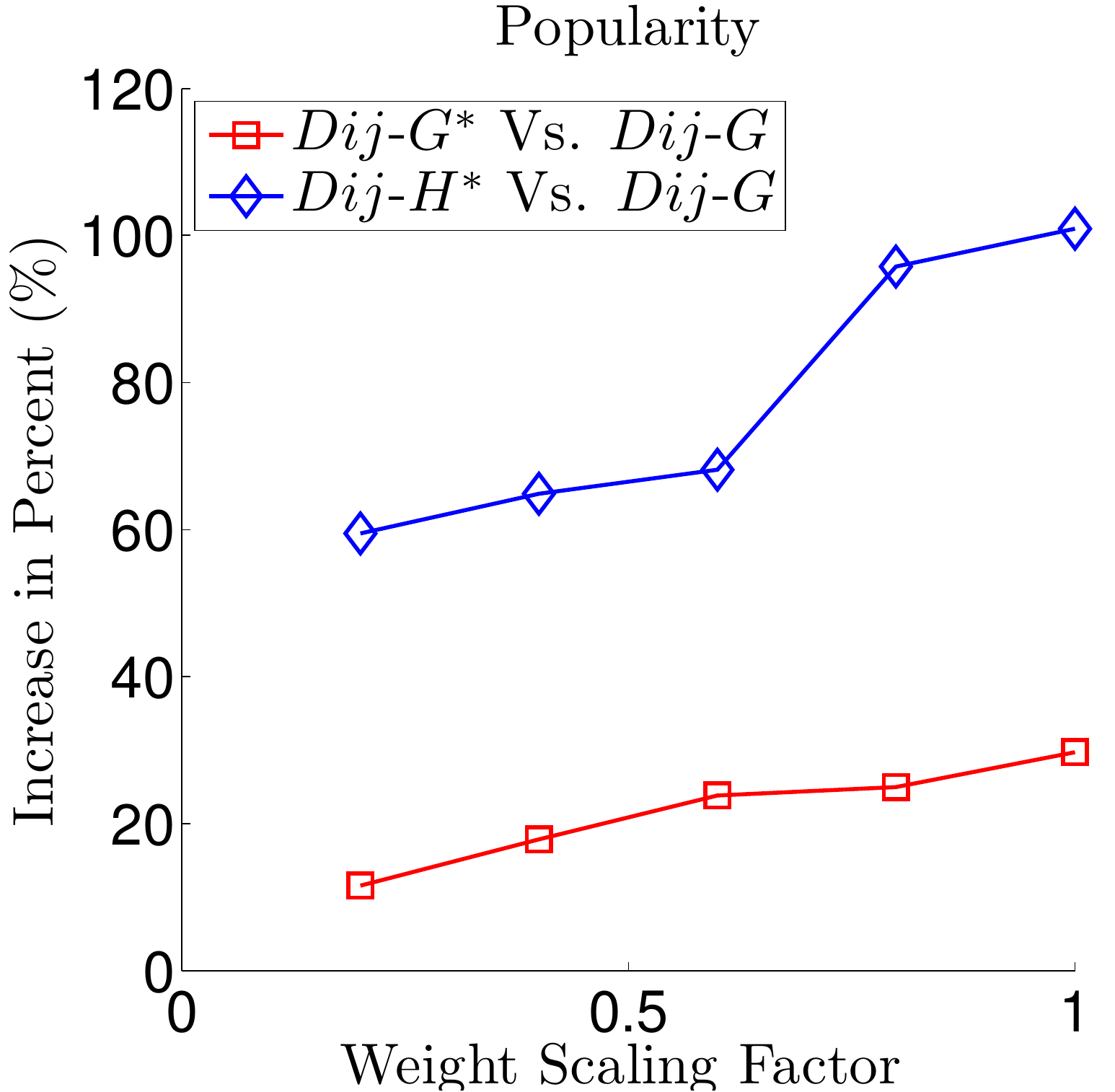}\label{subfig:LondonFlickrVsWeight}}
\subfigure[]{\includegraphics[width=0.245\textwidth]{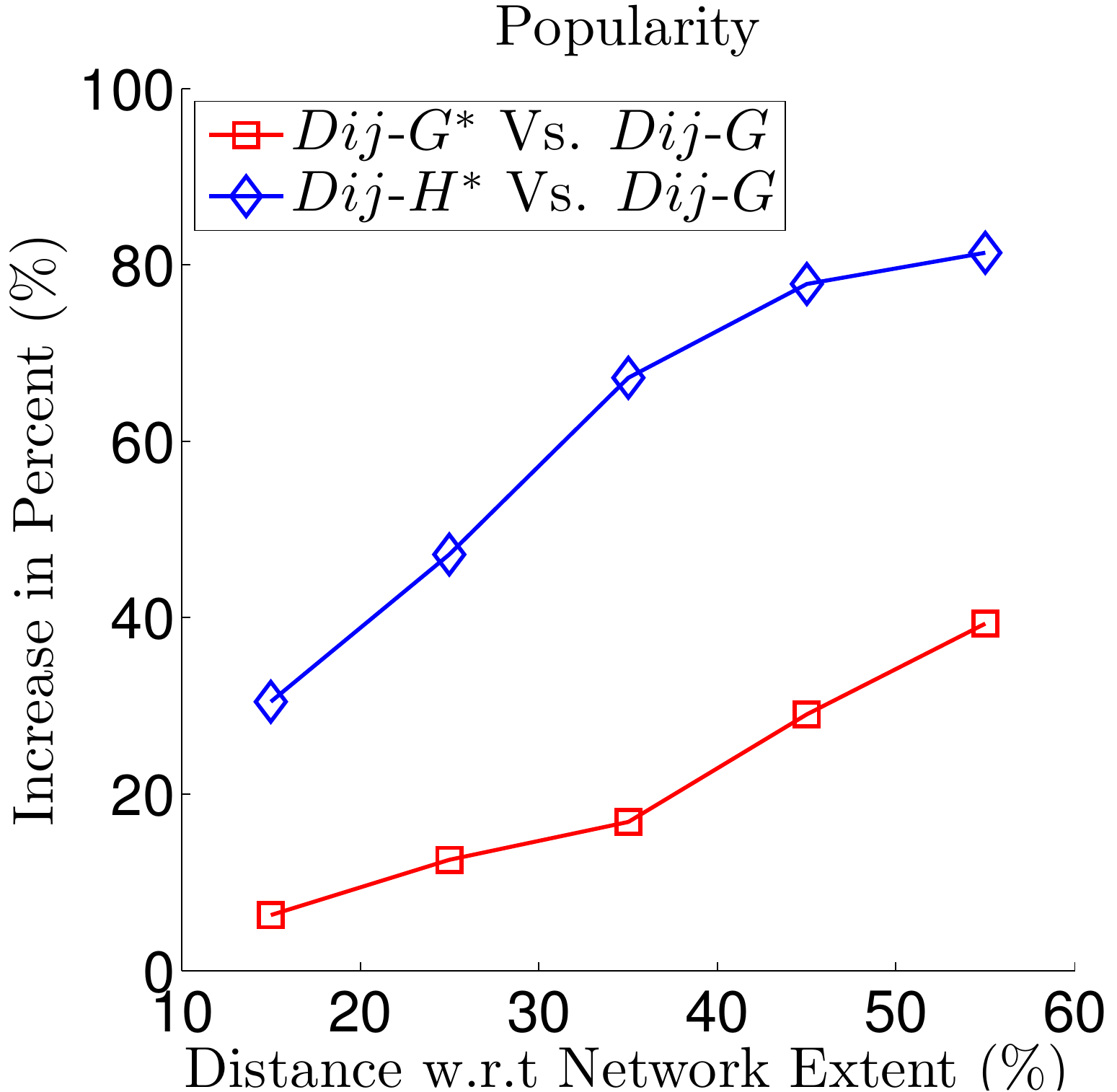}\label{subfig:LondonFlickrVsDistExt}}
\caption{(a), (c) show Distance and Flickr popularity increase for algorithms \gstard \ and \hstard \ for London dataset for experimental Setting $i$.
(b), (d) show Distance and Flickr popularity increase for algorithms \gstard \ and \hstard \ for London dataset for experimental Setting $ii$.}
\label{fig:london}
\end{figure}

In Setting $(i)$, for 100 randomly chosen pairs of start and target nodes the respective shortest
paths within the actual road network are computed using Dijkstra's algorithm, \gd. Continuing,
for the same start and target pairs, we run \gstard \ and \hstard. Subsequently, for each pair the
difference w.r.t.\ ER, distance and popularity is computed, and finally averaged
out over all pairs. We require the distance between start and target nodes to be at least 30\% and
at most 50\% of the Euclidean extent of the network (approximately 6km to 10km), in
order to obtain significant values of enrichment ratio and popularity and in order to exclude
paths which start and end in the outskirts of the city (where there are few to no PoIs).
Figure~\ref{fig:allER} ((a), (b), (c)) show the influence of the different scalings of $\wij$
on ER for the datasets of Paris, New York and London respectively. As we increase $\wij$ influence we observe an increase of ER for both \gstard \ and \hstard \ in comparison to \gd \ in all three datasets.
The increase in ER is in the range of 20\% to 200\% for the \gstard \ and in the range of 80\% to 700\% for \hstard.
Moreover, 
Figure~\ref{fig:paris}, Figure~\ref{fig:ny} and Figure~\ref{fig:london} ((a), (c))
show the influence of the different scalings of $\wij$ on distance and popularity. As we increase the influence of $\wij$ from 0.2 to 1.0,
we observe an increase of distance and popularity for both \gstard \ and \hstard \ in comparison to \gd \ in all three datasets.
The increase among all datasets, in terms of path length is in the range of 1\% to 15\% for \gstard \ and in the range of
5\% to 25\% for \hstard. Additionally, the increase in popularity is in the range of 10\% to 150\% for \gstard \ and in the range of 40\% to 200\% for \hstard.

It is clear that \gstard \ always performs better than \hstard \ in terms of path length increase, but \hstard \
performs always better in terms of ER and popularity. This is because \hstard \
routes directly through the PoIs, causing greater detours, but passing along highly weighted parts
of the enriched graph \gstar, which mostly coincide with dense Flickr regions.


Continuing, in Setting $(ii)$ the full weights are taken into account, but the proportionate distance of start and target node w.r.t.\ the extent
of the whole network is varied. We consider five different distance brackets of shortest path in the original graph \g, the first one ranging from 10\% to 20\%,
the last one ranging from 50\% to 60\% of the extent of the whole network. For 100 randomly chosen pairs of start and target nodes (within the respective distance bracket)
paths with \gd \ , \gstard \ and \hstard \ are computed. As before, for each pair the difference w.r.t.\ ER, distance and popularity is computed and averaged out over all pairs.
Figure~\ref{fig:allER} ((d), (e), (f)) show the increase of ER as we proceed through the distance brackets in all three datasets for the datasets of Paris, New York and London respectively.
The increase in ER is in the range of 10\% to 100\% for \gstard \ and in the range of 60\% to 700\% for \hstard.
Figure~\ref{fig:paris}, Figure~\ref{fig:ny} and Figure~\ref{fig:london} ((b), (d)) show the results:
As we proceed through the distance brackets, we observe an increase of the distance and popularity for \gstard \ as well as \hstard \ in comparison to \gd \ in all three datasets.
The increase among all datasets, in terms of path length, is in the range of 1\% to 15\% for \gstard \ and in the range of 5\% to 25\% for \hstard.
Finally, the increase in terms of popularity is in the range of 5\% to 60\% for \gstard \ and in the range of 30\% to 120\% for \hstard.
As in our previous experimental setting, it is clear that \gstard \ always outperforms \hstard \ in terms of path length increase, while \hstard \ always outperforms \gstard \ in terms of enrichment ratio and popularity.
The explanation for these results is, similarly with setting $i$, that \hstard \ routes directly through the PoIs and therefore accumulates greater ER and higher popularity values.

Overall, we may conclude that both \gstard \ and \hstard \ show convincing results. Both algorithms yield significant increase in terms of ER as well as in terms of the
independent Flickr-based measure popularity, while increasing path length only slightly. In the best case, ER increase amounts to 700\% and
200\% increase in terms of popularity (in comparison to the conventional shortest paths, as computed by \gd \ ), while the worst case increase in path length is only 25\%.
Consequently, we can claim that spatial relations, extracted from crowdsourced  information, can indeed be used to enrich actual road networks and define an alternative kind of routing which
reflects what people perceive as ``close''.

\begin{figure}[!h]
\begin{center}
\subfigure[Setting i]{\includegraphics[width=0.45\textwidth]{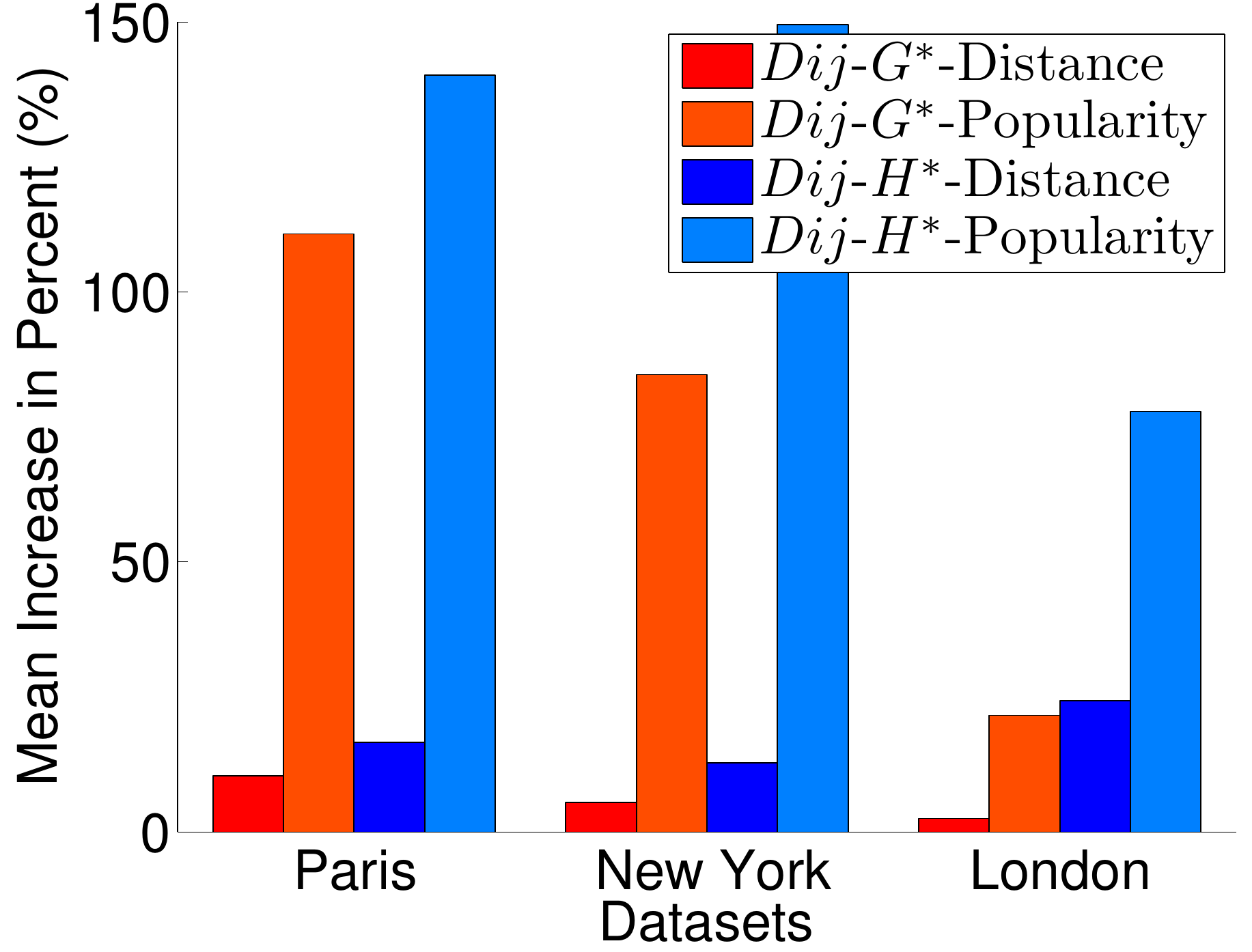}\label{subfig:NYDistVsWeight}}
\vspace{10pt}
\subfigure[Setting ii]{\includegraphics[width=0.45\textwidth]{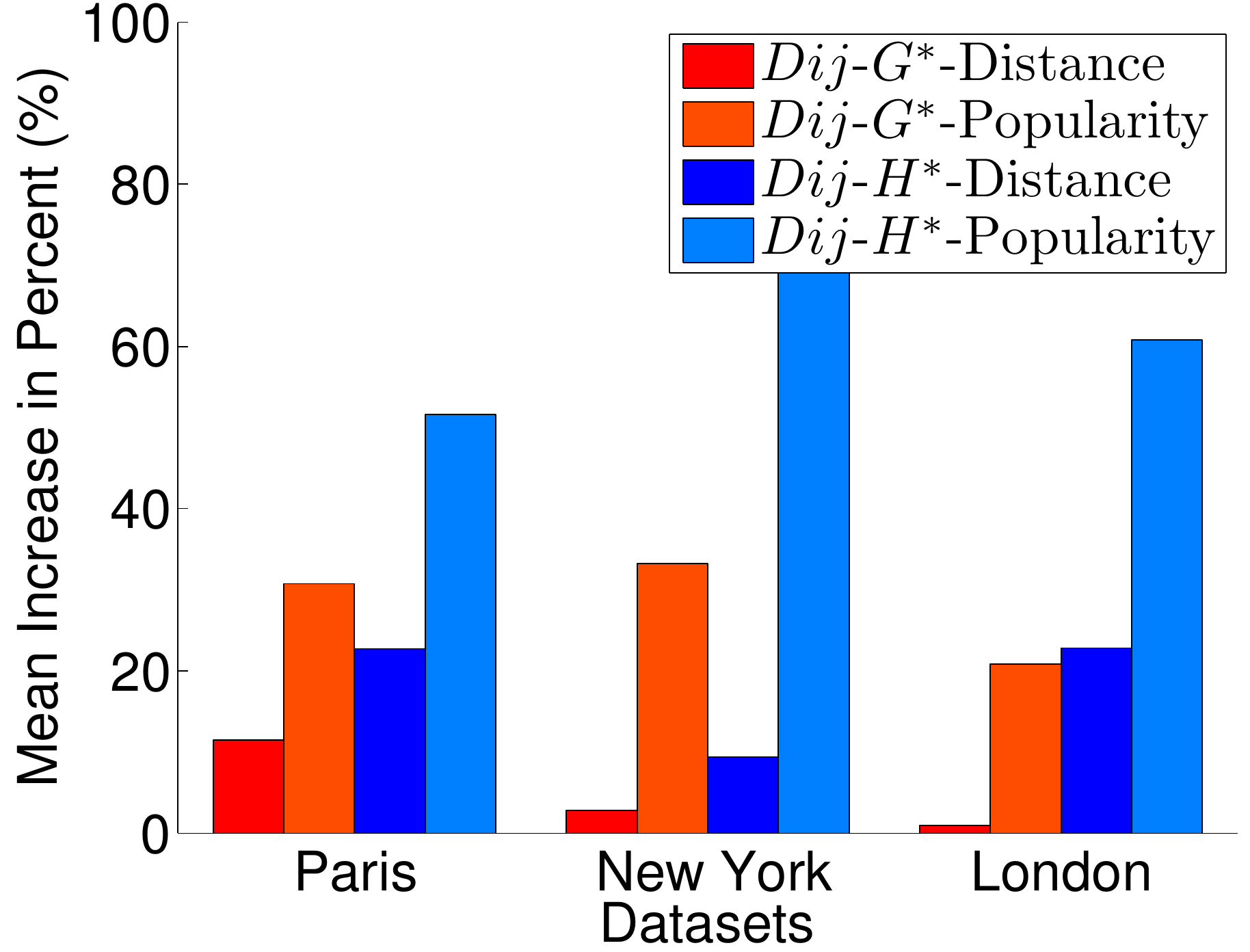}\label{subfig:NYDistVsDistExt}}
\end{center}
\caption{Trade-off between distance and popularity of paths.}
\label{fig:bars}
\end{figure}

Finally, Figure \ref{fig:bars} illustrates the trade-off that we take by deviating from the shortest path in order to obtain more interesting paths. This figure shows the relative increase 
in distance and popularity of the paths returned by our proposed approaches \gstard \ and \hstard, compared to the baseline approach \gd. 
For all three data sets, we can observe that using the \gstard \ approach we can obtain a significant increase in popularity of up to $150\%$ for the meager price of no more than $10\%$ additional distance incurred. 
In contrast, using the \hstard \ approach we get an even higher increase in popularity of the returned routes. However, this increase in popularity incurs a more significant distance overhead of up to $20\%$. 
In this experiments we used a scaling factor of $\alpha=0.5$. The trade-off between popularity and distance can be further balanced by adapting $\alpha$, as we have seen in Figures 8-10 ((a),(c)).

\vspace{-5pt}
\section{Conclusions and Outlook} 
\label{sec:conclusions}
\vspace{6pt}

In this work we presented an approach to computing knowledge-enriched paths within road networks.
We incorporated and developed novel methods to extract spatial relations between pairs of Points
of Interest such as ``near'' or ``close by'' from crowdsourced textual data, namely travel blogs.
We quantified the extracted relations using probabilistic models to handle the inherent
uncertainty of user-generated content. Based on these models, we proposed a new cost function to enrich real world road networks,
reflecting the closeness aspect according to the crowd. In contrast to existing approaches, we
did not enrich previously computed paths with semantical information, but the entire network.
Continuing, two routing algorithms were presented taking this closeness aspect into account.
Finally, we evaluated our ideas on three real world road network datasets, i.e., Paris, France,
New York City, USA, and London, UK. We used metadata from geotagged Flickr photos as a
ground truth to support our initial goal of providing more popular paths.

For future work, we are researching alternative methods for aggregating all categories of spatial relations.
Furthermore, we would like to investigate ways to suggest the popular path descriptions to the user
based on the Points of Interest they will encounter underway.

\vspace{10pt}


\bibliographystyle{abbrv}
\normalsize
\bibliography{geoinfo2}

\begin{thebibliography}{10}

\bibitem{Alvares:EnrichTraj}
L.~O. Alvares, V.~Bogorny, B.~Kuijpers, J.~A.~F. de~Macedo, B.~Moelans, and
  A.~Vaisman.
\newblock A model for enriching trajectories with semantic geographical
  information.
\newblock In {\em Proc. of the 15th Annual ACM Int'l Symp. on Advances in
  Geographic Information Systems}, pages 22:1--22:8, 2007.

\bibitem{Bishop}
C.~M. Bishop.
\newblock {\em Pattern Recognition and Machine Learning (Information Science
  and Statistics)}.
\newblock Springer-Verlag New York, Inc., Secaucus, NJ, USA, 2006.

\bibitem{Dempster}
A.~P. Dempster, N.~M. Laird, and D.~B. Rubin.
\newblock Maximum likelihood from incomplete data via the em algorithm.
\newblock {\em Journal of the Royal Statistical Society, Series B},
  39(1):1--38, 1977.

\bibitem{Kuhn:SIT}
M.~Duckham and L.~Kulik.
\newblock Simplest paths: Automated route selection for navigation.
\newblock In {\em Spatial Information Theory. Foundations of Geographic
  Information Science}, pages 169--185. 2003.

\bibitem{Feldman:iDiary}
D.~Feldman, A.~Sugaya, C.~Sung, and D.~Rus.
\newblock idiary: From gps signals to a text-searchable diary.
\newblock In {\em Proc. of the 11th ACM Conf. on Embedded Networked Sensor
  Systems}, pages 6:1--6:12, 2013.

\bibitem{GraKriRenSch11}
F.~Graf, H.-P. Kriegel, M.~Renz, and M.~Schubert.
\newblock Mario: Multi-attribute routing in open street map.
\newblock In {\em Advances in Spatial and Temporal Databases}, pages 486--490.
  2011.

\bibitem{Raubal:GIS}
S.~Hansen, K.-F. Richter, and A.~Klippel.
\newblock Landmarks in openls — a data structure for cognitive ergonomic
  route directions.
\newblock In M.~Raubal, H.~Miller, A.~Frank, and M.~Goodchild, editors, {\em
  Geographic Information Science}, volume 4197 of {\em Lecture Notes in
  Computer Science}, pages 128--144. Springer Berlin Heidelberg, 2006.

\bibitem{nltk}
E.~Loper and S.~Bird.
\newblock Nltk: The natural language toolkit.
\newblock In {\em Proc. of the ETMTNLP}, pages 63--70, 2002.

\bibitem{Lv:PersonalSemPlaceMining}
M.~Lv, L.~Chen, and G.~Chen.
\newblock Discovering personally semantic places from gps trajectories.
\newblock In {\em Proc. of the 21st ACM CIKM}, pages 1552--1556, 2012.

\bibitem{Flickr}
H.~Mousselly-Sergieh, D.~Watzinger, B.~Huber, M.~D\"{o}ller, E.~Egyed-Zsigmond,
  and H.~Kosch.
\newblock World-wide scale geotagged image dataset for automatic image
  annotation and reverse geotagging.
\newblock In {\em Proc. of the 5th ACM MMSys}, pages 47--52, 2014.

\bibitem{Palma:ClusteringApprForPlaceMining}
A.~T. Palma, V.~Bogorny, B.~Kuijpers, and L.~O. Alvares.
\newblock A clustering-based approach for discovering interesting places in
  trajectories.
\newblock In {\em Proc. of the ACM Symp. on Applied Computing}, pages 863--868,
  2008.

\bibitem{EPFL:SemanticTraj1}
C.~Parent, S.~Spaccapietra, C.~Renso, G.~Andrienko, N.~Andrienko, V.~Bogorny,
  M.~L. Damiani, A.~Gkoulalas-Divanis, J.~Macedo, N.~Pelekis, Y.~Theodoridis,
  and Z.~Yan.
\newblock Semantic trajectories modeling and analysis.
\newblock {\em ACM Comput. Surv. '13}, 45(4):42:1--42:32, Aug. 2013.

\bibitem{DBLP:Yahoo}
D.~Quercia, R.~Schifanella, and L.~M. Aiello.
\newblock The shortest path to happiness: Recommending beautiful, quiet, and
  happy routes in the city.
\newblock {\em CoRR}, abs/1407.1031, 2014.

\bibitem{Raubal:2002:EWI:646933.759822}
M.~Raubal and S.~Winter.
\newblock Enriching wayfinding instructions with local landmarks.
\newblock In {\em Proceedings of the Second International Conference on
  Geographic Information Science}, GIScience '02, pages 243--259, London, UK,
  UK, 2002. Springer-Verlag.

\bibitem{Sacharidis}
D.~Sacharidis and P.~Bouros.
\newblock Routing directions: Keeping it fast and simple.
\newblock In {\em Proc. of the 21st ACM SIGSPATIAL GIS}, pages 164--173, 2013.

\bibitem{skoumas}
G.~Skoumas, D.~Pfoser, and A.~Kyrillidis.
\newblock On quantifying qualitative geospatial data: A probabilistic approach.
\newblock In {\em Proc. of the 2nd ACM GEOCROWD}, pages 71--78, 2013.

\bibitem{EPFL:SemanticTraj4}
S.~Spaccapietra and C.~Parent.
\newblock Adding meaning to your steps.
\newblock In {\em Proc. of the 30th Int'l Conf. on Conceptual Modeling}, pages
  13--31, 2011.

\bibitem{Verbeek}
J.~J. Verbeek, N.~Vlassis, and B.~Kröse.
\newblock Efficient greedy learning of gaussian mixture models.
\newblock {\em Neural Computation}, 15:469--485, 2003.

\bibitem{Westphal:QualitativeRouting}
M.~Westphal and J.~Renz.
\newblock Evaluating and minimizing ambiguities in qualitative route
  instructions.
\newblock In {\em Proc. of the 19th ACM Int'l Conf. on Advances in Geographic
  Information Systems}, pages 171--180, 2011.

\bibitem{Yan:SemanticAnnotation}
Z.~Yan, D.~Chakraborty, C.~Parent, S.~Spaccapietra, and K.~Aberer.
\newblock Semitri: A framework for semantic annotation of heterogeneous
  trajectories.
\newblock In {\em Proc. of the 14th Int'l Conf. on Extending Database
  Technology}, pages 259--270, 2011.

\bibitem{EPFL:SemanticTraj2}
Z.~Yan, D.~Chakraborty, C.~Parent, S.~Spaccapietra, and K.~Aberer.
\newblock Semantic trajectories: Mobility data computation and annotation.
\newblock {\em ACM Trans. Intell. Syst. Technol.}, 4(3):49:1--49:38, July 2013.

\bibitem{EPFL:SemanticTraj6}
Z.~Yan, L.~Spremic, D.~Chakraborty, C.~Parent, S.~Spaccapietra, and K.~Aberer.
\newblock Automatic construction and multi-level visualization of semantic
  trajectories.
\newblock In {\em Proc. of the 18th Int'l Conf. on Advances in Geographic
  Information Systems}, pages 524--525, 2010.

\end{thebibliography}
\section{Appendix} 
\label{sec:appendix}

The re-estimation of parameters $\lambda_i = \{w_i, \mu_i, \Sigma_i\}$ of a given GMM $p(d|\lambda)$ per each EM step, can be accomplished by the application of the following equations.

For all components $i \in \{1, . . . , M\}$ \newline \newline

\begin{equation}
	P(i|D_j) = \frac{w_i g(D_j;\lambda_i)}{p(D_j|\lambda)} 
\end{equation}

\begin{equation}
	w_i = \sum_{j=1}^{n} \frac{P(i|D_j)}{n} 
\end{equation}

\begin{equation}
	\mu_i = \sum_{j=1}^{n} \frac{P(i|D_j)D_j}{n w_i} 
\end{equation}

\begin{equation}
	\Sigma_{i} = \sum_{j=1}^{n} \frac{P(i|D_j)(D_j - \mu_i)(D_j-\mu_i)^{\intercal}}{n w_{i}} 
\end{equation}

EM re-estimates model parameters $\lambda = (\lambda_1,\dots,\lambda_M)$ until the convergence of log-likelihood $\mathcal{L}$ given by Equation~\ref{eq:loglike}. 
As mentioned before, the EM algorithm is a greedy parameter estimation algorithm and is not guaranteed to lead us to the solution yielding maximum log-likelihood on 
$\mathcal{D}$ among all maxima of the log-likelihood. Nevertheless, using the EM algorithm, if we are ``close'' to the global optimum (maximum) 
of the parameter space, then it is very likely we can obtain the globally optimal solution.

\end{document}